\documentclass[11pt]{article}%
\usepackage{amsfonts}
\usepackage{amssymb}
\usepackage{amsmath}
\usepackage{graphicx}
\usepackage{amscd}%
\providecommand{\U}[1]{\protect\rule{.1in}{.1in}}

\newtheorem{theorem}{Theorem}

\newtheorem{proposition}[theorem]{Proposition}

\begin{document}

\title{Information as Distinctions:\\New Foundations for Information Theory}
\author{David Ellerman\\University of California/Riverside}
\maketitle

\begin{abstract}
The logical basis for information theory is the newly developed logic of
partitions that is dual to the usual Boolean logic of subsets. The key concept
is a "distinction" of a partition, an ordered pair of elements in distinct
blocks of the partition. The logical concept of entropy based on partition
logic is the normalized counting measure of the set of distinctions of a
partition on a finite set--just as the usual logical notion of probability
based on the Boolean logic of subsets is the normalized counting measure of
the subsets (events). Thus logical entropy is a measure on the set of ordered
pairs, and all the compound notions of entropy (join entropy, conditional
entropy, and mutual information) arise in the usual way from the measure
(e.g., the inclusion-exclusion principle)--just like the corresponding notions
of probability. The usual Shannon entropy of a partition is developed by
replacing the normalized count of distinctions (dits) by the average number of
binary partitions (bits) necessary to make all the distinctions of the partition.

\end{abstract}
\tableofcontents

\section{Introduction}

Information is about making distinctions or differences. In James Gleick's
book, \textit{The Information: A History, A Theory, A Flood}, he noted the
focus on differences in the seventeenth century polymath, John Wilkins, who
was a founder of the Royal Society. In 1641, the year before Newton was born,
Wilkins published one of the earliest books on cryptography, \textit{Mercury
or the Secret and Swift Messenger}, which not only pointed out the fundamental
role of differences but noted that any (finite) set of different things could
be encoded by words in a binary code.

\begin{quotation}
\noindent For in the general we must note, That whatever is capable of a
competent Difference, perceptible to any Sense, may be a sufficient Means
whereby to express the Cogitations. It is more convenient, indeed, that these
Differences should be of as great Variety as the Letters of the Alphabet; but
it is sufficient if they be but twofold, because Two alone may, with somewhat
more Labour and Time, be well enough contrived to express all the rest.
\cite[Chap. XVII, p. 69]{wilkins:merc}
\end{quotation}

\noindent Wilkins explains that a five letter binary code would be sufficient
to code the letters of the alphabet since $2^{5}=32$.

\begin{quotation}
\noindent Thus any two Letters or Numbers, suppose A.B. being transposed
through five Places, will yield Thirty Two Differences, and so consequently
will superabundantly serve for the Four and twenty Letters... .\cite[Chap.
XVII, p. 69]{wilkins:merc}
\end{quotation}

\noindent As Gleick noted:

\begin{quotation}
\noindent Any difference meant a binary choice. Any binary choice began the
expressing of cogitations. Here, in this arcane and anonymous treatise of
1641, the essential idea of information theory poked to the surface of human
thought, saw its shadow, and disappeared again for [three] hundred years.
\cite[p. 161]{gleick:info}
\end{quotation}

In this paper, we will start afresh by deriving an information-as-distinctions
notion of logical entropy \cite{ell:countdits} from the new logic of
partitions \cite{ell:partitions} that is mathematically dual to the usual
Boolean logic of subsets. Then the usual Shannon entropy \cite{shannon:comm}
will be essentially derived from the concepts behind logical entropy as
another way to measure information-as-distinctions. This treatment of the
various notions of Shannon entropy (e.g., mutual, conditional, and joint
entropy) will also explain why their interrelations can be represented using a
Venn diagram picture \cite{camp:meas}.

\section{Logical Entropy}

\subsection{Partition logic}

The logic normally called "propositional logic" is a special case of the logic
of subsets originally developed by George Boole \cite{boole:lot}. In the
Boolean logic of subsets of a fixed non-empty universe set $U$, the variables
in formulas refer to subsets $S\subseteq U$ and the logical operations such as
the join $S\vee T$, meet $S\wedge T$, and implication $S\Rightarrow T$ are
interpreted as the subset operations of union $S\cup T$, intersection $S\cap
T$, and the conditional $S\Rightarrow T=S^{c}\cup T$. Then "propositional"
logic is the special case where $U=1$ is the one-element set whose subsets
$\emptyset$ and $1$ are interpreted as the truth values $0$ and $1$ (or false
and true) for propositions.

In subset logic, a \textit{valid formula} or \textit{tautology} is a formula
such as $\left[  S\wedge\left(  S\Rightarrow T\right)  \right]  \Rightarrow T$
where for any non-empty $U$, no matter what subsets of $U$ are substituted for
the variables, the whole formula evaluates to $U$ by the subset operations. It
is a theorem that if a formula is valid just for the special case of $U=1$
(i.e., as in a truth table tautology), then it is valid for any $U$. But in
today's textbook treatments of so-called "propositional" logic, the
truth-table version of a tautology is usually given as a definition, not as a
theorem in subset logic.

What is lost by restricting attention to the special case of propositional
logic rather than the general case of subset logic? At least two things are
lost, and both are relevant for our development.

\begin{itemize}
\item Firstly if it is developed as the logic of subsets, then it is natural,
as Boole did, to attach a quantitative measure to each subset $S$ of a finite
universe $U$, namely the normalized counting measure $\frac{|S|}{\left\vert
U\right\vert }$ which can be interpreted as the \textit{logical probability}
$\Pr\left(  S\right)  $ (where the elements of $U$ are assumed equiprobable)
of randomly drawing an element from $S$.

\item Secondly, the notion of a subset (unlike the notion of a proposition)
has a mathematical dual in the notion of a quotient set, as is evidenced by
the dual interplay between subobjects (subgroups, subrings,...) and quotient
objects throughout abstract algebra.
\end{itemize}

This duality is the "turn-around-the-arrows" category-theoretic duality, e.g.,
between monomorphisms and epimorphisms, applied to sets \cite{law:sfm}. The
notion of a quotient set of $U$ is equivalent to the notion of an equivalence
relation on $U$ or a partition $\pi=\left\{  B\right\}  $ of $U$. When Boole's
logic is seen as the logic of subsets (rather than propositions), then the
notion arises of a dual logic of partitions which has now been developed
\cite{ell:partitions}.

\subsection{Logical Entropy}

A partition $\pi=\left\{  B\right\}  $ on a finite set $U$ is a set of
non-empty disjoint subsets $B$ ("blocks" of the partition) of $U$ whose union
is $U$. The idea of information-as-distinctions is made precise by defining a
\textit{distinction} or \textit{dit} \textit{of a partition} $\pi=\left\{
B\right\}  $ of $U$ as an ordered pair $\left(  u,u^{\prime}\right)  $ of
elements $u,u^{\prime}\in U$ that are in different blocks of the partition.
The notion of "a distinction of a partition" plays the analogous role in
partition logic as the notion of "an element of a subset" in subset logic. The
set of distinctions of a partition $\pi$ is its \textit{dit set}
$\operatorname*{dit}\left(  \pi\right)  $. The subsets of $U$ are partially
ordered by inclusion with the universe set $U$ as the top of the order and the
empty set $\emptyset$ as the bottom of the order. A partition $\pi=\left\{
B\right\}  $ \textit{refines} a partition $\sigma=\left\{  C\right\}  $,
written $\sigma\preceq\pi$, if each block $B\in\pi$ is contained in some block
$C\in\sigma$. The partitions of $U$ are partially ordered by refinement which
is equivalent to the inclusion ordering of dit sets. The discrete partition
$\mathbf{1=}\left\{  \left\{  u\right\}  \right\}  _{u\in U}$, where the
blocks are all the singletons, is the top of the order, and the indiscrete
partition $\mathbf{0=}\left\{  U\right\}  $ (with just one block $U$) is the
bottom. Only the self-pairs $\left(  u,u\right)  \in\Delta\subseteq U\times U$
of the diagonal $\Delta$ can never be a distinction. All the possible
distinctions $U\times U-\Delta$ are the dits of $\mathbf{1}$ and no dits are
distinctions of $\mathbf{0}$ just as all the elements are in $U$ and none in
$\emptyset$.

In this manner, we can construct a table of analogies between subset logic and
partition logic.

\begin{center}%
\begin{tabular}
[c]{|c||c|c|}\hline
& Subset logic & Partition logic\\\hline\hline
`Elements' & Elements $u$ of $S$ & Dits $\left(  u,u^{\prime}\right)  $ of
$\pi$\\\hline
Order & Inclusion $S\subseteq T$ & Refinement: $\operatorname*{dit}\left(
\sigma\right)  \subseteq\operatorname*{dit}\left(  \pi\right)  $\\\hline
Top of order & $U$ all elements & $\operatorname*{dit}(\mathbf{1)}%
=U^{2}-\Delta$, all dits\\\hline
Bottom of order & $\emptyset$ no elements & $\operatorname*{dit}%
(\mathbf{0)=\emptyset}$, no dits\\\hline
Variables in formulas & Subsets $S$ of $U$ & Partitions $\pi$ on $U$\\\hline
Operations & Subset ops. & Partition ops. \cite{ell:partitions}\\\hline
Formula $\Phi(x,y,...)$ holds & $u$ element of $\Phi(S,T,...)$ & $\left(
u,u^{\prime}\right)  $ dit of $\Phi(\pi,\sigma,...)$\\\hline
Valid formula & $\Phi(S,T,...)=U$, $\forall S,T,...$ & $\Phi(\pi
,\sigma,...)=\mathbf{1}$, $\forall\pi,\sigma,...$\\\hline
\end{tabular}

Table of analogies between subset and partition logics
\end{center}

A dit set $\operatorname*{dit}\left(  \pi\right)  $ of a partition on $U$ is a
subset of $U\times U$ of a particular kind, namely the complement of an
equivalence relation. An \textit{equivalence relation} is reflexive,
symmetric, and transitive. Hence the complement is a subset $P\subseteq
U\times U$ that is:

\begin{enumerate}
\item irreflexive (or anti-reflexive), $P\cap\Delta=\emptyset$;

\item symmetric, $\left(  u,u^{\prime}\right)  \in P$ implies $\left(
u^{\prime},u\right)  \in P$; and

\item anti-transitive (or co-transitive), if $\left(  u,u^{\prime\prime
}\right)  \in P$ then for any $u^{\prime}\in U$, $\left(  u,u^{\prime}\right)
\in P$ or $\left(  u^{\prime},u^{\prime\prime}\right)  \in P$,
\end{enumerate}

\noindent and such binary relations will be called \textit{partition relations
}(also called \textit{apartness relations}).

Given any subset $S\subseteq U\times U$, the
\textit{reflexive-symmetric-transitive (rst) closure} $\overline{S^{c}}$ of
the complement $S^{c}$ is the smallest equivalence relation containing $S^{c}%
$, so its complement is the largest partition relation contained in $S$, which
is called the \textit{interior} $\operatorname*{int}\left(  S\right)  $ of
$S$. This usage is consistent with calling the subsets that equal their
rst-closures \textit{closed subsets} of $U\times U$ (so closed subsets =
equivalence relations) so the complements are the \textit{open subsets} (=
partition relations). However it should be noted that the rst-closure is not a
topological closure since the closure of a union is not necessarily the union
of the closures, so the "open" subsets do not form a topology on $U\times U$.

The interior operation $\operatorname*{int}:\wp\left(  U\times U\right)
\rightarrow\wp\left(  U\times U\right)  $ provides a universal way to define
operations on partitions from the corresponding subset operations:

\begin{quote}
apply the subset operation to the dit sets and then, if necessary, take the
interior to obtain the dit set of the partition operation.
\end{quote}

\noindent Given partitions $\pi=\left\{  B\right\}  $ and $\sigma=\left\{
C\right\}  $ on $U$, their \textit{join} $\pi\vee\sigma$ is the partition
whose dit set $\operatorname*{dit}\left(  \pi\vee\sigma\right)  $ is the
interior of $\operatorname*{dit}\left(  \pi\right)  \cup\operatorname*{dit}%
\left(  \sigma\right)  $ (since the union $\cup$ is the subset join
operation). But the union of partition relations (open subsets) is a partition
relation (open subset) so that:

\begin{center}
$\operatorname*{dit}\left(  \pi\vee\sigma\right)  =\operatorname*{dit}\left(
\pi\right)  \cup\operatorname*{dit}\left(  \sigma\right)  $.
\end{center}

\noindent This gives the same join $\pi\vee\sigma$ as the usual definition
which is the partition whose blocks are the non-empty intersections $B\cap C$
for $B\in\pi$ and $C\in\sigma$. To define the \textit{meet} $\pi\wedge\sigma$
of the two partitions, we apply the subset meet operation of intersection to
the dit sets and then take the interior (which is necessary in this case):

\begin{center}
$\operatorname*{dit}\left(  \pi\wedge\sigma\right)  =\operatorname*{int}%
\left[  \operatorname*{dit}\left(  \pi\right)  \cap\operatorname*{dit}\left(
\sigma\right)  \right]  $.
\end{center}

\noindent This gives the same result as the usual definition of the partition
meet in the literature.\footnote{But note that many authors think in terms of
equivalence relations instead of partition relations and thus reverse the
definitions of the join and meet. Hence their "lattice of partitions" is
really the lattice of equivalence relations, the opposite of the partition
lattice $\Pi\left(  U\right)  $ defined here with refinement as the ordering
relation.} Perhaps surprisingly, the other logical operations such as the
implication do not seem to be defined for partitions in the literature. Since
the subset operation of implication is $S\Rightarrow T=S^{c}\cup T$, we define
the \textit{partition implication} $\sigma\Rightarrow\pi$ as the partition
whose dit set is:

\begin{center}
$\operatorname*{dit}\left(  \pi\Rightarrow\sigma\right)  =int\left[
\operatorname*{dit}\left(  \sigma\right)  ^{c}\cup\operatorname*{dit}\left(
\pi\right)  \right]  $.\footnote{The equivalent but more perspicuous
definition of $\sigma\Rightarrow\pi$ is the partition that is like $\pi$
except that whenever a block $B\in\pi$ is contained in a block $C\in\sigma$,
then $B$ is `discretized' in the sense of being replaced by all the singletons
$\left\{  u\right\}  $ for $u\in B$. Then it is immediate that the refinement
$\sigma\preceq\pi$ holds iff $\sigma\Rightarrow\pi=\mathbf{1}$, as we would
expect from the corresponding relation, $S\subseteq T$ iff $S\Rightarrow
T=S^{c}\cup T=U$, in subset logic.}
\end{center}

\noindent The refinement partial order $\sigma\preceq\pi$ is just inclusion of
dit sets, i.e., $\sigma\preceq\pi$ iff $\operatorname*{dit}\left(
\sigma\right)  \subseteq\operatorname*{dit}\left(  \pi\right)  $. If we denote
the lattice of partitions (using the refinement ordering) as $\Pi\left(
U\right)  $, then the mapping:

\begin{center}
$\operatorname*{dit}:\Pi\left(  U\right)  \rightarrow\wp\left(  U\times
U\right)  $

Dit set representation of partition lattice
\end{center}

\noindent represents the lattice of partitions as the lattice $\mathcal{O}%
\left(  U\times U\right)  $ of open subsets (under inclusion) of $\wp\left(
U\times U\right)  $.

For any finite set $X$, a (finite) \textit{measure} $\mu$ is a function
$\mu:\wp\left(  X\right)  \rightarrow%
\mathbb{R}
$ such that:

\begin{enumerate}
\item $\mu\left(  \emptyset\right)  =0$,

\item for any $E\subseteq X$, $\mu\left(  E\right)  \geq0$, and

\item for any disjoint subsets $E_{1}$ and $E_{2}$, $\mu(E_{1}\cup E_{2}%
)=\mu\left(  E_{1}\right)  +\mu\left(  E_{2}\right)  $.
\end{enumerate}

Any finite set $X$ has the \textit{counting measure} $\left\vert \ \right\vert
:\wp\left(  X\right)  \rightarrow%
\mathbb{R}
$ and \textit{normalized counting measure} $\frac{\left\vert \ \right\vert
}{\left\vert X\right\vert }:\wp\left(  X\right)  \rightarrow%
\mathbb{R}
$ defined on the subsets of $X$. Hence for finite $U$, we have the counting
measure $\left\vert \ \right\vert $ and the normalized counting measure
$\frac{\left\vert \ \right\vert }{\left\vert U\times U\right\vert }$ defined
on $\wp\left(  U\times U\right)  $. Boole used the normalized counting measure
$\frac{\left\vert \ \right\vert }{\left\vert U\right\vert }$ defined on the
power-set Boolean algebra $\wp\left(  U\right)  $ to define the logical
probability $\Pr\left(  S\right)  =\frac{\left\vert S\right\vert }{\left\vert
U\right\vert }$ of an event $S\subseteq U$.\cite{boole:lot} \noindent In view
of the analogy between elements in subset logic and dits in partition logic,
the construction analogous to the logical probability is the normalized
counting measure applied to dit sets. That is the definition of the:

\begin{center}
$h\left(  \pi\right)  =\frac{\left\vert \operatorname*{dit}\left(  \pi\right)
\right\vert }{\left\vert U\times U\right\vert }$

\textit{Logical entropy of a partition} $\pi$.
\end{center}

\noindent Thus the logical entropy function $h\left(  {}\right)  $ is the dit
set representation composed with the normalized counting measure:

\begin{center}
$h:\Pi\left(  U\right)  \rightarrow%
\mathbb{R}
=\Pi\left(  U\right)  \overset{\operatorname*{dit}}{\longrightarrow
}\mathcal{\wp}\left(  U\times U\right)  \overset{\frac{\left\vert
\ \right\vert }{\left\vert U\times U\right\vert }}{\longrightarrow}%
\mathbb{R}
$.

Logical entropy function
\end{center}

One immediate consequence is the inclusion-exclusion principle:

\begin{center}
$\frac{\left\vert \operatorname*{dit}\left(  \pi\right)  \cap
\operatorname*{dit}\left(  \sigma\right)  \right\vert }{\left\vert U\times
U\right\vert }=\frac{\left\vert \operatorname*{dit}\left(  \pi\right)
\right\vert }{\left\vert U\times U\right\vert }+\frac{\left\vert
\operatorname*{dit}\left(  \sigma\right)  \right\vert }{\left\vert U\times
U\right\vert }-\frac{\left\vert \operatorname*{dit}\left(  \pi\right)
\cup\operatorname*{dit}\left(  \sigma\right)  \right\vert }{\left\vert U\times
U\right\vert }=h\left(  \pi\right)  +h\left(  \sigma\right)  -h\left(  \pi
\vee\sigma\right)  $
\end{center}

\noindent which provides the motivation for our definition below of
$\frac{\left\vert \operatorname*{dit}\left(  \pi\right)  \cap
\operatorname*{dit}\left(  \sigma\right)  \right\vert }{\left\vert U\times
U\right\vert }$ as the "logical mutual information" of the partitions $\pi$
and $\sigma$.

In a random (i.e., equiprobable) drawing of an element from $U$, the event $S
$ occurs with the probability $\Pr\left(  S\right)  $. If we take two
independent (i.e., with replacement) random drawings from $U$, i.e., pick a
random ordered pair from $U\times U$, then $h\left(  \pi\right)  $ is the
probability that the pair is a distinction of $\pi$, i.e., that $\pi$
distinguishes. These analogies are summarized in the following table which
uses the language of probability theory (e.g., set of outcomes, events, the
occurrence of an event):

\begin{center}%
\begin{tabular}
[c]{|c||c|c|}\hline
& Subset logic & Partition logic\\\hline\hline
`Outcomes' & Elements $u$ of $S$ & Ordered pairs $\left(  u,u^{\prime}\right)
\in U\times U$\\\hline
`Events' & Subsets $S$ of $U$ & Partitions $\pi$ of $U$\\\hline
`Event occurs' & $u\in S$ & $\left(  u,u^{\prime}\right)  \in
\operatorname*{dit}\left(  \pi\right)  $\\\hline
Norm. counting measure & $\Pr\left(  S\right)  =\frac{\left\vert S\right\vert
}{\left\vert U\right\vert }$ & $h\left(  \pi\right)  =\frac{\left\vert
\operatorname*{dit}\left(  \pi\right)  \right\vert }{\left\vert U\times
U\right\vert }$\\\hline
Interpretation & Prob. event $S$ occurs & Prob. partition $\pi$
distinguishes\\\hline
\end{tabular}

Table of quantitative analogies between subset and partition logics.
\end{center}

Thus logical entropy $h(\pi)$ is the simple quantitative measure of the
distinctions of a partition $\pi$ just as the logical probability $\Pr\left(
S\right)  $ is the quantitative measure of the elements in a subset $S$. In
short, information theory is to partition logic as probability theory is to
ordinary subset logic.

To generalize logical entropy from partitions to finite probability
distributions, note that:

\begin{center}
$\operatorname*{dit}(\pi)=\left\{  B\times B^{\prime}:B,B^{\prime}\in\pi,B\neq
B^{\prime}\right\}  =U\times U-\left\{  B\times B:B\in\pi\right\}  $.
\end{center}

\noindent Using $p_{B}=\frac{\left\vert B\right\vert }{\left\vert U\right\vert
}$, we have:

\begin{center}
$h\left(  \pi\right)  =\frac{\left\vert \operatorname*{dit}\left(  \pi\right)
\right\vert }{\left\vert U\times U\right\vert }=\frac{|U|^{2}-\sum_{B\in\pi
}|B|^{2}}{\left\vert U\right\vert ^{2}}=1-\sum_{B\in\pi}\left(  \frac
{|B|}{\left\vert U\right\vert }\right)  ^{2}=1-\sum_{B\in\pi}p_{B}^{2}$.
\end{center}

\noindent An ordered pair $\left(  u,u^{\prime}\right)  \in B\times B$ for
some $B\in\pi$ is an \textit{indistinction} or \textit{indit} of $\pi$ where
$\operatorname*{indit}\left(  \pi\right)  =U\times U-\operatorname*{dit}%
\left(  \pi\right)  $. Hence in a random drawing of a pair from $U\times U$,
$\sum_{B\in\pi}p_{B}^{2}$ is the probability of drawing an indistinction,
while $h\left(  \pi\right)  =1-\sum_{B\in\pi}p_{B}^{2}$ is the probability of
drawing a distinction.

Entropies will be defined both for partitions on finite sets and for finite
probability distributions (i.e., finite random variables). Given a random
variable $u$ with the probability distribution $p=\left(  p_{1},...,p_{n}%
\right)  $ over the $n$ distinct values $U=\left\{  u_{1},...,u_{n}\right\}
$, a distinction of the discrete partition on $U$ is just a pair $\left(
u_{i},u_{j}\right)  $ with $i\neq j$ and with the probability $p_{i}p_{j}$.
Applying the previous notion to the logical entropy of a partition to this
case with $p_{B}=p_{i}$ (where $B=\left\{  u_{i}\right\}  $), we have the:

\begin{center}
$h\left(  p\right)  =1-\sum_{i}p_{i}^{2}=\sum_{i}p_{i}\left(  1-p_{i}\right)
$

\textit{Logical entropy of a finite probability distribution} $p$%
.\footnote{This could be taken as the logical entropy $h(u)$ of the random
variable $u$ but since the values of $u$ are irrelevant (other than being
distinct for $i\neq j$), we can take the logical entropy $h\left(  p\right)  $
as a function solely of the probability distribution $p$ of the random
variable.}
\end{center}

Since $1=\left(  \sum_{i=1}^{n}p_{i}\right)  ^{2}=\sum_{i}p_{i}^{2}%
+\sum_{i\neq j}p_{i}p_{j}$, we again have the logical entropy $h\left(
p\right)  $ as the probability $\sum_{i\neq j}p_{i}p_{j}$ of drawing a
distinction in two independent samplings of the probability distribution $p$.
This is also clear from defining the product measure on the subsets
$S\subseteq U\times U$:

\begin{center}
$\mu\left(  S\right)  =\sum\left\{  p_{i}p_{j}:\left(  u_{i},u_{j}\right)  \in
S\right\}  $

\textit{Product measure} on $U\times U$
\end{center}

\noindent Then the logical entropy $h\left(  p\right)  =\mu\left(
\mathbf{1}_{U}\right)  $ is just the product measure of the$\ $dit set of the
discrete partition on $U$. There is also the obvious generalization to
consider any partition $\pi$ on $U$ and then define for each block $B\in\pi$,
$p_{B}=\sum_{u_{i}\in B}p_{i}$. Then the logical entropy $h\left(  \pi\right)
=\mu\left(  \operatorname*{dit}\left(  \pi\right)  \right)  $ is the product
measure of the dit set of $\pi$ (so it is still interpreted as the probability
of drawing a distinction of $\pi$) and that is equivalent to $\sum_{B}%
p_{B}\left(  1-p_{B}\right)  $.

For the uniform distribution $p_{i}=\frac{1}{n}$, the logical entropy has its
maximum value of $1-\frac{1}{n}$. Regardless of the first draw (even for a
different probability distribution over the same $n$ outcomes), the
probability that the second draw is different is $1-\frac{1}{n}$. The logical
entropy has its minimum value of $0$ for $p=\left(  1,0,...,0\right)  $ so that:

\begin{center}
$0\leq h\left(  p\right)  \leq1-\frac{1}{n}$.
\end{center}

An important special case is a set $U$ of $\left\vert U\right\vert =N$
equiprobable elements and a partition $\pi$ on $U$ with $n$ equal-sized blocks
of $N/n$ elements each. Then the number of distinctions of elements is
$N^{2}-n\left(  \frac{N}{n}\right)  ^{2}=N^{2}-\frac{N^{2}}{n}$ which
normalizes to the logical entropy of $h\left(  \pi\right)  =1-\frac{1}{n}$ and
which is independent of $N$. Thus it holds when $N=n$ and we take the elements
to be the equal blocks themselves. Thus for an equal-blocked partition on a
set of equiprobable elements, the normalized number of distinctions of
elements is the same as the normalized number of distinctions of blocks, and
that quantity is the:

\begin{center}
$h\left(  p_{0}\right)  =1-p_{0}=1-\frac{1}{n}$

\textit{Logical entropy of an equiprobable set of }$n$\textit{\ elements}.
\end{center}

\subsection{A statistical treatment of logical entropy}

It might be noted that no averaging is involved in the interpretation of
$h\left(  \pi\right)  $. It is the number of distinctions $\left\vert
\operatorname*{dit}\left(  \pi\right)  \right\vert $ normalized. The
definition of the logical entropy $h\left(  p\right)  =\sum_{i=1}^{n}%
p_{i}h\left(  p_{i}\right)  =\sum_{i=1}^{n}p_{i}\left(  1-p_{i}\right)  $ of a
probability distribution $p=\left(  p_{1},...,p_{n}\right)  $ is in the form
of the average value of the random variable which has the value $h\left(
p_{i}\right)  =1-p_{i}$ with the probability $p_{i}$.

Hence the formula can be arrived at by applying the law of large numbers in
the form where the finite random variable $X$ takes the value $x_{i}$ with
probability $p_{i}$:

\begin{center}
$\lim_{N\rightarrow\infty}\frac{1}{N}\sum_{j=1}^{N}x_{j}=\sum_{i=1}^{n}%
p_{i}x_{i}$.
\end{center}

At each step $j$ in repeated independent sampling $u_{1}u_{2}...u_{N}$ of the
probability distribution $p=\left(  p_{1},...,p_{n}\right)  $, the probability
that the $j^{th}$ result $u_{j}$ was \textit{not} $u_{j}$ is $1-\Pr\left(
u_{j}\right)  $ so the \textit{average} probability of the result being
different than it was at each place in that sequence is:

\begin{center}
$\frac{1}{N}\sum_{j=1}^{N}\left(  1-\Pr\left(  u_{j}\right)  \right)  $.
\end{center}

In the long run, the \textit{typical} sequences will dominate where the
$i^{th}$ outcome is sampled $p_{i}N$ times so that we have the value $1-p_{i}$
occurring $p_{i}N$ times:

\begin{center}
$\lim_{N\rightarrow\infty}\frac{1}{N}\sum_{j=1}^{N}\left(  1-\Pr\left(
u_{j}\right)  \right)  =\frac{1}{N}\sum_{i=1}^{n}p_{i}N\left(  1-p_{i}\right)
=h\left(  p\right)  $.
\end{center}

The logical entropy $h\left(  p\right)  =\sum_{i}p_{i}\left(  1-p_{i}\right)
=\sum_{i\neq j}p_{i}p_{j}$ is usually interpreted as the \textit{pair-drawing
probability of getting distinct outcomes} from the distribution $p=\left(
p_{1},...,p_{n}\right)  $. Now we have a different interpretation of logical
entropy as \textit{the average probability of being different}.

\subsection{A brief history of the logical entropy formula}

The logical entropy formula $h\left(  p\right)  =\sum_{i}p_{i}\left(
1-p_{i}\right)  =1-\sum_{i}p_{i}^{2}$ is the probability of getting distinct
values $u_{i}\neq u_{j}$ in two independent samplings of the random variable
$u$. The complementary measure $1-h\left(  p\right)  =\sum_{i}p_{i}^{2}$ is
the probability that the two drawings yield the same value from $U$. Thus
$1-\sum_{i}p_{i}^{2}$ is a measure of heterogeneity or diversity in keeping
with our theme of information as distinctions, while the complementary measure
$\sum_{i}p_{i}^{2}$ is a measure of homogeneity or concentration.
Historically, the formula can be found in either form depending on the
particular context. The $p_{i}$'s might be relative shares such as the
relative share of organisms of the $i^{th}$ species in some population of
organisms, and then the interpretation of $p_{i}$ as a probability arises by
considering the random choice of an organism from the population.

According to I. J. Good, the formula has a certain naturalness:

\begin{quote}
If $p_{1},...,p_{t}$ are the probabilities of $t$ mutually exclusive and
exhaustive events, any statistician of this century who wanted a measure of
homogeneity would have take about two seconds to suggest $\sum p_{i}^{2}$
which I shall call $\rho$.\ \cite[p. 561]{good:div}
\end{quote}

As noted by Bhargava and Uppuluri \cite{bhar:gini}, the formula $1-\sum
p_{i}^{2}$ was used by Gini in 1912 (\cite{gini:vem} reprinted in \cite[p.
369]{gini:vemrpt}) as a measure of \textquotedblleft
mutability\textquotedblright\ or diversity. But another development of the
formula (in the complementary form) in the early twentieth century was in
cryptography. The American cryptologist, William F. Friedman, devoted a 1922
book (\cite{fried:ioc}) to the "index of coincidence"\ (i.e., $\sum p_{i}^{2}
$). Solomon Kullback (of the Kullback-Leibler divergence treated later) worked
as an assistant to Friedman and wrote a book on cryptology which used the
index. \cite{kull:crypt}

During World War II, Alan M. Turing worked for a time in the Government Code
and Cypher School at the Bletchley Park facility in England. Probably unaware
of the earlier work, Turing used $\rho=\sum p_{i}^{2}$ in his cryptoanalysis
work and called it the \textit{repeat rate} since it is the probability of a
repeat in a pair of independent draws from a population with those
probabilities (i.e., the identification probability $1-h\left(  p\right)  $).
Polish cryptoanalyists had independently used the repeat rate in their work on
the Enigma \cite{rej:polish}.

After the war, Edward H. Simpson, a British statistician, proposed $\sum
_{B\in\pi}p_{B}^{2}$ as a measure of species concentration (the opposite of
diversity) where $\pi$ is the partition of animals or plants according to
species and where each animal or plant is considered as equiprobable. And
Simpson gave the interpretation of this homogeneity measure as "the
probability that two individuals chosen at random and independently from the
population will be found to belong to the same group."\cite[p. 688]{simp:md}
Hence $1-\sum_{B\in\pi}p_{B}^{2}$ is the probability that a random ordered
pair will belong to different species, i.e., will be distinguished by the
species partition. In the biodiversity literature \cite{ric:unify}, the
formula is known as "Simpson's index of diversity"\ or sometimes, the
\textit{Gini-Simpson index \cite{rao:div}}.\ However, Simpson along with I. J.
Good worked at Bletchley Park during WWII, and, according to Good, "E. H.
Simpson and I both obtained the notion [the repeat rate] from
Turing."\ \cite[p. 395]{good:turing} When Simpson published the index in 1948,
he (again, according to Good) did not acknowledge Turing "fearing that to
acknowledge him would be regarded as a breach of security."\ \cite[p.
562]{good:div}

In 1945, Albert O. Hirschman (\cite[p. 159]{hirsch:np} and \cite{hirsch:pat})
suggested using $\sqrt{\sum p_{i}^{2}}$ as an index of trade concentration
(where $p_{i}$ is the relative share of trade in a certain commodity or with a
certain partner). A few years later, Orris Herfindahl \cite{her:conc}
independently suggested using $\sum p_{i}^{2}$ as an index of industrial
concentration (where $p_{i}$ is the relative share of the $i^{th}$ firm in an
industry). In the industrial economics literature, the index $H=\sum p_{i}%
^{2}$ is variously called the Hirschman-Herfindahl index, the HH index, or
just the H index of concentration. If all the relative shares were equal
(i.e., $p_{i}=1/n$), then the identification or repeat probability is just the
probability of drawing any element, i.e., $H=1/n$, so $\frac{1}{H}=n$ is the
number of equal elements. This led to the "numbers equivalent"\ interpretation
of the reciprocal of the H index \cite{adel:ne}. In general, given an event
with probability $p_{0}$, the \textit{numbers-equivalent\ interpretation} of
the event is that it is `as if' an element was drawn out of a set $U_{1/p_{0}%
}$ of $\frac{1}{p_{0}}$ equiprobable elements (it is `as if' since $1/p_{0}$
need not be an integer). This interpretation will be used later in the dit-bit connection.

In view of the frequent and independent discovery and rediscovery of the
formula $\rho=\sum p_{i}^{2}$ or its complement $1-\sum p_{i}^{2}$ by Gini,
Friedman, Turing, Hirschman, Herfindahl, and no doubt others, I. J. Good
wisely advises that "it is unjust to associate $\rho$ with any one
person."\ \cite[p. 562]{good:div}

Two elements from $U=\left\{  u_{1},...,u_{n}\right\}  $ are either identical
or distinct. Gini \cite{gini:vem} introduced $d_{ij}$ as the \textquotedblleft
distance\textquotedblright\ between the $i^{th}$ and $j^{th}$ elements where
$d_{ij}=1$ for $i\not =j$ and $d_{ii}=0$. Since $1=\left(  p_{1}%
+...+p_{n}\right)  \left(  p_{1}+...+p_{n}\right)  =\sum_{i}p_{i}^{2}%
+\sum_{i\not =j}p_{i}p_{j}$, the logical entropy, i.e., Gini's index of
mutability, $h\left(  p\right)  =1-\sum_{i}p_{i}^{2}=\sum_{i\not =j}p_{i}%
p_{j}$, is the average logical distance between a pair of independently drawn
elements. But one might generalize by allowing other distances $d_{ij}=d_{ji}$
for $i\not =j$ (but always $d_{ii}=0$) so that $Q=\sum_{i\not =j}d_{ij}%
p_{i}p_{j}$ would be the average distance between a pair of independently
drawn elements from $U$. In 1982, C. R. (Calyampudi Radhakrishna) Rao
introduced precisely this concept as \textit{quadratic entropy} \cite{rao:div}%
. In many domains, it is quite reasonable to move beyond the bare-bones
\textit{logical distance} of $d_{ij}=1$ for $i\not =j$ (i.e., the complement
$1-\delta_{ij}$ of the Kronecker delta) so that Rao's quadratic entropy is a
useful and easily interpreted generalization of logical
entropy.\footnote{Rao's treatment also includes (and generalizes) the natural
extension to continuous (square-integrable) probability density functions
$f\left(  x\right)  $: $h\left(  f\right)  =1-\int f\left(  x\right)  ^{2}%
dx$.}

\section{Shannon Entropy}

\subsection{Shannon-Hartley entropy of a set}

The Shannon entropy will first be motivated in the usual fashion and then
developed from the basic logical notion of entropy. Shannon, like Ralph
Hartley \cite{hart:ti} before him, starts with the question of how much
"information" is required to single out a designated element from a set $U$ of
equiprobable elements. This is often formulated in terms of the search
\cite{Renyi:pt} for a hidden element like the answer in a Twenty Questions
game or the sent message in a communication. But being able to always find the
designated element is equivalent to being able to distinguish all elements
from one another. That is, if the designated element was in a set of two or
more elements that had not been distinguished from one another, then one would
not be able to single out the designated element. Thus "singling out" or
"identifying" an element in a set is just another way to conceptualize
"distinguishing" all the elements of the set.

Intuitively, one might measure "information" as the minimum number of
yes-or-no questions in a game of Twenty Questions that it would take in
general to \textit{distinguish} all the possible "answers" (or "messages" in
the context of communications). This is readily seen in the simple case where
$\left\vert U\right\vert =2^{m}$, i.e., the size of the set of equiprobable
elements is a power of $2$. Then following the lead of Wilkins over three
centuries earlier, the $2^{m}$ elements could be encoded using words of length
$m$ in a binary code such as the digits $\left\{  0,1\right\}  $ of binary
arithmetic (or $\left\{  A,B\right\}  $ in the case of Wilkins). Then an
efficient or minimum set of yes-or-no questions needed to single out the
hidden element is the set of $m$ questions:

\begin{center}
"Is the $j^{th}$ digit in the binary code for the hidden element a $1$?"
\end{center}

\noindent for $j=1,...,m$. Each element is distinguished from any other
element by their binary codes differing in at least one digit. The information
gained in finding the outcome of an equiprobable binary trial, like flipping a
fair coin, is what Shannon calls a \textit{bit} (derived from "binary digit").
Hence the information gained in distinguishing all the elements out of $2^{m}$
equiprobable elements is:

\begin{center}
$m=\log_{2}\left(  2^{m}\right)  =\log_{2}\left(  \left\vert U\right\vert
\right)  =\log_{2}\left(  \frac{1}{p_{0}}\right)  $ bits
\end{center}

\noindent where $p_{0}=\frac{1}{2^{m}}$ is the probability of any given
element (henceforth all logs to base $2$).

This is usefully restated in terms of partitions. Given two partitions
$\pi=\left\{  B\right\}  $ and $\sigma=\left\{  C\right\}  $ of $U$, their
\textit{join} $\pi\vee\sigma$ is the partition of $U$ whose blocks are the
non-empty intersections $B\cap C$ for $B\in\pi$ and $C\in\sigma$. The
determination of the $j^{th}$ digit in the binary code for the hidden element
defines a binary partition $\pi_{j}$ of $U$. Then to say that the answers to
the $m$ questions above distinguish all the elements means that the join, $%
{\textstyle\bigvee\nolimits_{j=1}^{m}}
\pi_{j}=\mathbf{1}$, is the discrete partition on the set $U$ with cardinality
$2^{m}$. Thus we could also take $m=\log\left(  \frac{1}{p_{0}}\right)  $ as
the minimum number of binary partitions necessary to distinguish the elements
(i.e., to single out any given element).

In the more general case where $\left\vert U\right\vert =n$ is not a power of
$2$, we extrapolate to the definition of $H\left(  p_{0}\right)  $ where
$p_{0}=\frac{1}{n}$ as:

\begin{center}
$H\left(  p_{0}\right)  =\log\left(  \frac{1}{p_{0}}\right)  =\log\left(
n\right)  $

\textit{Shannon-Hartley entropy for an equiprobable set} $U$ of $n$ elements.
\end{center}

\noindent The definition is further extrapolated to the case where we are only
given a probability $p_{0}$ so that we say that $H\left(  p_{0}\right)
=\log\left(  \frac{1}{p_{0}}\right)  $ binary partitions are needed to
distinguish a set of $\frac{1}{p_{0}}$ elements when $\frac{1}{p_{0}}$ is not
an integer.

\subsection{Shannon entropy of a probability distribution}

This interpretation of the special case of $2^{m}$ or more generally $1/p_{0}
$ equiprobable elements is extended to an arbitrary finite probability
distribution $p=\left(  p_{1},...,p_{n}\right)  $ by an averaging process. For
the $i^{th}$ outcome ($i=1,...,n$), its probability $p_{i}$ is "as if" it were
drawn from a set of $\frac{1}{p_{i}}$ equiprobable elements (ignoring that
$\frac{1}{p_{i}}$ may not be an integer for this averaging argument) so the
Shannon-Hartley information content of distinguishing the equiprobable
elements of such a set would be $\log\left(  \frac{1}{p_{i}}\right)  $. But
that occurs with probability $p_{i}$ so the probabilistic average gives the
usual definition of the:

\begin{center}
$H\left(  p\right)  =\sum_{i=1}^{n}p_{i}H\left(  p_{i}\right)  =\sum_{i=1}%
^{n}p_{i}\log\left(  \frac{1}{p_{i}}\right)  =-\sum_{i=1}^{n}p_{i}\log\left(
p_{i}\right)  $

\textit{Shannon entropy of a finite probability distribution} $p$.
\end{center}

For the uniform distribution $p_{i}=\frac{1}{n}$, the Shannon entropy has it
maximum value of $\log\left(  n\right)  $ while the minimum value is $0$ for
the trivial distribution $p=(1,0,...,0)$ so that:

\begin{center}
$0\leq H\left(  p\right)  \leq\log\left(  n\right)  $.
\end{center}

\subsection{A statistical treatment of Shannon entropy}

Shannon makes this averaging argument rigorous by using the law of large
numbers. Suppose that we have a three-letter alphabet $\left\{  a,b,c\right\}
$ where each letter was equiprobable, $p_{a}=p_{b}=p_{c}=\frac{1}{3}$, in a
multi-letter message. Then a one-letter or two-letter message cannot be
exactly coded with a binary $0,1$ code with equiprobable $0$'s and $1$'s. But
any probability can be better and better approximated by longer and longer
representations in the binary number system. Hence we can consider longer and
longer messages of $N$ letters along with better and better approximations
with binary codes. The long run behavior of messages $u_{1}u_{2}...u_{N}$
where $u_{i}\in\left\{  a,b,c\right\}  $ is modeled by the law of large
numbers so that the letter $a$ will tend to occur $p_{a}N=\frac{1}{3}N$ times
and similarly for $b$ and $c$. Such a message is called \textit{typical}.

The probability of any one of those typical messages is:

\begin{center}
$p_{a}^{p_{a}N}p_{b}^{p_{b}N}p_{c}^{p_{c}N}=\left[  p_{a}^{p_{a}}p_{b}^{p_{b}%
}p_{c}^{p_{c}}\right]  ^{N}$
\end{center}

or, in this case,

\begin{center}
$\left[  \left(  \frac{1}{3}\right)  ^{1/3}\left(  \frac{1}{3}\right)
^{1/3}\left(  \frac{1}{3}\right)  ^{1/3}\right]  ^{N}=\left(  \frac{1}%
{3}\right)  ^{N}$.
\end{center}

\noindent Hence the number of such typical messages is $3^{N}$.

If each message was assigned a unique binary code, then the number of $0,1$'s
in the code would have to be $X$ where $2^{X}=3^{N}$ or $X=\log\left(
3^{N}\right)  =N\log\left(  3\right)  $. Hence the number of equiprobable
binary questions or bits needed per letter of the messages is:

\begin{center}
$N\log(3)/N=\log\left(  3\right)  =3\times\frac{1}{3}\log\left(  \frac{1}%
{1/3}\right)  =H\left(  p\right)  $.
\end{center}

\noindent This example shows the general pattern.

In the general case, let $p=\left(  p_{1},...,p_{n}\right)  $ be the
probabilities over a $n$-letter alphabet $A=\left\{  a_{1},...,a_{n}\right\}
$. In an $N$-letter message, the probability of a particular message
$u_{1}u_{2}...u_{N}$ is $\Pi_{i=1}^{N}\Pr\left(  u_{i}\right)  $ where $u_{i}$
could be any of the symbols in the alphabet so if $u_{i}=a_{j}$ then
$\Pr\left(  u_{i}\right)  =p_{j}$.

In a \textit{typical} message, the $i^{th}$ symbol will occur $p_{i}N$ times
(law of large numbers) so the probability of a typical message is (note change
of indices to the letters of the alphabet):

\begin{center}
$\Pi_{k=1}^{n}p_{k}^{p_{k}N}=\left[  \Pi_{k=1}^{n}p_{k}^{p_{k}}\right]  ^{N}$.
\end{center}

Since the probability of a typical message is $P^{N}$ for $P=\Pi_{k=1}%
^{n}p_{k}^{p_{k}}$, the typical messages are equiprobable. Hence the number of
typical messages is $\left[  \Pi_{k=1}^{n}p_{k}^{-p_{k}}\right]  ^{N}$ and
assigning a unique binary code to each typical message requires $X$ bits where
$2^{X}=\left[  \Pi_{k=1}^{n}p_{k}^{-p_{k}}\right]  ^{N}$ where:

\begin{center}
$X=\log\left\{  \left[  \Pi_{k=1}^{n}p_{k}^{-p_{k}}\right]  ^{N}\right\}
=N\log\left[  \Pi_{k=1}^{n}p_{k}^{-p_{k}}\right]  $

$=N\sum_{k=1}^{n}\log\left(  p_{k}^{-p_{k}}\right)  =N\sum_{k}-p_{k}%
\log\left(  p_{k}\right)  $

$=N\sum_{k}p_{k}\log\left(  \frac{1}{p_{k}}\right)  =NH\left(  p\right)  $.
\end{center}

Hence the Shannon entropy $H\left(  p\right)  =\sum_{k=1}^{n}p_{k}\log\left(
\frac{1}{p_{k}}\right)  $ is interpreted as the limiting \textit{average
number of bits necessary per letter in the message}. In terms of distinctions,
this is the \textit{average number of binary partitions necessary per letter
to distinguish the messages}. It is this averaging result that allows us to
consider "the number of binary partitions it takes to distinguish the elements
of $U$" when $\left\vert U\right\vert $ is not a power of $2$ since "number"
is interpreted as "average number."

\subsection{Shannon entropy of a partition}

Shannon entropy can also be defined for a partition $\pi=\left\{  B\right\}  $
on a set $U$. If the elements of $U$ are equiprobable, then the probability
that a randomly drawn element is in a block $B\in\pi$ is $p_{B}=\frac
{\left\vert B\right\vert }{\left\vert U\right\vert }$. In a set of $\frac
{1}{p_{B}}$ equiprobable elements, it would take (on average) $H\left(
p_{B}\right)  =\log\left(  \frac{1}{p_{B}}\right)  $ binary partitions to
distinguish the elements. Averaging over the blocks, we have the:

\begin{center}
$H\left(  \pi\right)  =\sum_{B\in\pi}p_{B}\log\left(  \frac{1}{p_{B}}\right)
$

\textit{Shannon entropy of a partition} $\pi$.
\end{center}

\subsection{Shannon entropy and statistical mechanics}

The functional form of Shannon's formula is often further "justified" or
"motivated" by asserting that it is the same as the notion of entropy in
statistical mechanics, and hence the name "entropy." The name "entropy" is
here to stay but the justification of the formula by reference to statistical
mechanics is not quite correct. The connection between entropy in statistical
mechanics and Shannon's entropy is only via a numerical approximation, the
Stirling approximation, where if the first two terms in the Stirling
approximation are used, then the Shannon formula is obtained.

The first two terms in the Stirling approximation for $\ln(N!)$ are:
$\ln\left(  N!\right)  \approx N\ln(N)-N$. The first three terms in the
Stirling approximation are: $\ln\left(  N!\right)  \approx N(\ln
(N)-1)+\frac{1}{2}\ln\left(  2\pi N\right)  $.

If we consider a partition on a finite $U$ with $\left\vert U\right\vert =N$,
with $n$ blocks of size $N_{1},...,N_{n}$, then the number of ways of
distributing the individuals in these $n$ boxes with those numbers $N_{i}$ in
the $i^{th}$ box is: $W=\frac{N!}{N_{1}!\times...\times N_{n}!}$. The
normalized natural log of $W$, $S=\frac{1}{N}\ln\left(  W\right)  $ is one
form of entropy in statistical mechanics. Indeed, the formula "$S=k\log\left(
W\right)  $" is engraved on Boltzmann's tombstone.

The entropy formula can then be developed using the first two terms in the
Stirling approximation.

\begin{center}
$S=\frac{1}{N}\ln\left(  W\right)  =\frac{1}{N}\ln\left(  \frac{N!}%
{N_{1}!\times...\times N_{n}!}\right)  =\frac{1}{N}\left[  \ln(N!)-\sum_{i}%
\ln(N_{i}!)\right]  $

$\approx\frac{1}{N}\left[  N\left[  \ln\left(  N\right)  -1\right]  -\sum
_{i}N_{i}\left[  \ln\left(  N_{i}\right)  -1\right]  \right]  $

$=\frac{1}{N}\left[  N\ln(N)-\sum N_{i}\ln(N_{i})\right]  =\frac{1}{N}\left[
\sum N_{i}\ln\left(  N\right)  -\sum N_{i}\ln\left(  N_{i}\right)  \right]  $

$=\sum\frac{N_{i}}{N}\ln\left(  \frac{1}{N_{i}/N}\right)  =\sum p_{i}%
\ln\left(  \frac{1}{p_{i}}\right)  =H_{e}\left(  p\right)  $
\end{center}

\noindent where $p_{i}=\frac{N_{i}}{N}$ (and where the formula with logs to
the base $e$ only differs from the usual base $2$ formula by a scaling
factor). Shannon's entropy $H_{e}\left(  p\right)  $ is in fact an excellent
numerical approximation to $S=\frac{1}{N}\ln\left(  W\right)  $ for large $N$
(e.g., in statistical mechanics).

But the common claim is that Shannon's entropy has the \textit{same functional
form} as entropy in statistical mechanics, and that is simply false. If we use
a three-term Stirling approximation, then we obtain an even better numerical
approximation:\footnote{For the case $n=2$, MacKay \cite[p. 2]{mackay:info}
also uses Stirling's approximation to give a "more accurate approximation"
(using the next term in the Stirling approximation) to the entropy of
statistical mechanics than the Shannon entropy.}

\begin{center}
$S=\frac{1}{N}\ln\left(  W\right)  \approx H_{e}\left(  p\right)  +\frac
{1}{2N}\ln\left(  \frac{2\pi N^{n}}{\left(  2\pi\right)  ^{n}\Pi p_{i}%
}\right)  $
\end{center}

\noindent but no one would suggest using that "more accurate" entropy formula
in information theory. Shannon's formula should be justified and understood by
the arguments given previously, and not by over-interpreting the approximate
relationship with entropy in statistical mechanics.

\subsection{The basic dit-bit connection}

The basic datum is "the" set $U_{n}$ of $n$ elements with the equal
probabilities $p_{0}=\frac{1}{n}$. In that basic case of an equiprobable set,
we can derive the dit-bit connection, and then by using a probabilistic
average, we can develop the Shannon entropy, expressed in terms of bits, from
the logical entropy, expressed in terms of (normalized) dits, or vice-versa.

Given $U_{n}$ with $n$ equiprobable elements, the number of dits (of the
discrete partition on $U_{n}$) is $n^{2}-n$ so the normalized dit count is:

\begin{center}
$h\left(  p_{0}\right)  =h\left(  \frac{1}{n}\right)  =1-p_{0}=1-\frac{1}{n}$
normalized dits.
\end{center}

\noindent That is the dit-count or logical measure of the information is a set
of $n$ distinct elements.\footnote{The context will determine whether
"dit-count" refers to the "raw" count $\left\vert \operatorname*{dit}\left(
\pi\right)  \right\vert $ or the normalized count $\frac{\left\vert
\operatorname*{dit}\left(  \pi\right)  \right\vert }{\left\vert U\times
U\right\vert }$.}

But we can also measure the information in the set by the number of binary
partitions it takes (on average) to distinguish the elements, and that
bit-count is:

\begin{center}
$H\left(  p_{0}\right)  =H\left(  \frac{1}{n}\right)  =\log\left(  \frac
{1}{p_{0}}\right)  =\log\left(  n\right)  $ bits.
\end{center}

By solving the dit-count and the bit-count for $p_{0}$ and equating, we can
derive each measure in terms of the other:

\begin{center}
$H\left(  p_{0}\right)  =\log\left(  \frac{1}{1-h\left(  p_{0}\right)
}\right)  $ and $h\left(  p_{0}\right)  =1-\frac{1}{2^{H\left(  p_{0}\right)
}}$

The dit-bit conversion formulas.
\end{center}

The common thing being measured is an equiprobable $U_{n}$ where $n=\frac
{1}{p_{0}}$. The dit-count for $U_{n}$ is $h\left(  p_{0}\right)  =1-\frac
{1}{n}$ and the bit-count for $U_{n}$ is $H\left(  p_{0}\right)  =\log\left(
\frac{1}{p_{0}}\right)  $, and the bit-dit connection gives the relationship
between the two counts. Using this dit-bit connection between the two
different ways to measure the "information" in $U_{n}$, each entropy can be
developed from the other.

We start with the logical entropy of a probability distribution $p=\left(
p_{1},...,p_{n}\right)  $: $h\left(  p\right)  =\sum_{i=1}^{n}p_{i}h\left(
p_{i}\right)  $. It is expressed as the probabilistic average of the
dit-counts or logical entropies of the sets $U_{1/p_{i}}$ with $\frac{1}%
{p_{i}}$ equiprobable elements.\footnote{Starting with the datum of the
probability $p_{i}$, there is no necessity that $n=\frac{1}{p_{i}}$ is an
integer so the dit-counts for $U_{1/p_{i}}$ are extrapolations while the
bit-counts or binary partition counts for $U_{n} $ are already extrapolations
even when $n$ is an integer but not a power of $2$.} But if we switch to the
binary-partition bit-counts of the information content of those same sets
$U_{1/p_{i}}$ of $\frac{1}{p_{i}}$ equiprobable elements, then the bit-counts
are $H\left(  p_{i}\right)  =\log\left(  \frac{1}{p_{i}}\right)  $ and the
probabilistic average is the Shannon entropy: $H\left(  p\right)  =\sum
_{i=1}^{n}p_{i}H\left(  p_{i}\right)  $. Both entropies have the mathematical form:

\begin{center}
$\sum_{i}p_{i}\left(  \text{measure of info. in }U_{1/p_{i}}\right)  $
\end{center}

\noindent and differ by using either the dit-count or bit-count to measure the
information in $U_{1/p_{i}}$.

Clearly the process is reversible, so one can use the dit-bit connection in
reverse to develop the logical entropy $h\left(  p\right)  $ from the Shannon
entropy $H\left(  p\right)  $. Thus the two notions of entropy are simply two
different ways, using distinctions (dit-counts) or binary partitions
(bit-counts), to measure the information in a probability distribution.

Moreover the dit-bit connection carries over to the compound notions of
entropy so that the Shannon notions of conditional entropy, mutual
information, and joint entropy can be developed from the corresponding notions
for logical entropy. Since the logical notions are the values of a probability
measure, the compound notions of logical entropy have the usual Venn diagram
relations such as the inclusion-exclusion principle. There is a well-known
analogy between the "Venn diagram" relationships for the Shannon entropies and
the relationships satisfied by any measure on a set (\cite{abramson:it},
\cite{camp:meas}). As L. L. Campbell puts it, the analogy:

\begin{quotation}
\noindent suggests the possibility that $H\left(  \alpha\right)  $ and
$H\left(  \beta\right)  $ are measures of sets, that $H\left(  \alpha
,\beta\right)  $ is the measure of their union, that $I\left(  \alpha
,\beta\right)  $ is the measure of their intersection, and that $H\left(
\alpha|\beta\right)  $ is the measure of their difference. The possibility
that $I\left(  \alpha,\beta\right)  $ is the entropy of the "intersection" of
two partitions is particularly interesting. This "intersection," if it
existed, would presumably contain the information common to the partitions
$\alpha$ and $\beta$.\cite[p. 113]{camp:meas}
\end{quotation}

All of Campbell's desiderata are precisely true when:

\begin{itemize}
\item "sets" = dit sets, and

\item "entropies" = normalized counting measure of the (dit) sets, i.e., the
logical entropies.
\end{itemize}

Since the logical entropies are the values of a measure, by developing the
corresponding Shannon notions from the logical ones, we have an explanation of
why the Shannon notions also exhibit the same Venn diagram relationships.

The expository strategy is to first develop the Shannon and logical compound
notions of entropy separately and then to show the relationship using the
dit-bit connection.

\section{Conditional entropies}

\subsection{Logical conditional entropy}

Given two partitions $\pi=\left\{  B\right\}  $ and $\sigma=\left\{
C\right\}  $ on a finite set $U$, how might one measure the new information
that is provided by $\pi$ that was not already in $\sigma$? Campbell suggests
associating sets with partitions so the conditional entropy would be the
measure of the difference between the sets. Taking the information as
distinctions, we take the difference between the$\ $dit sets, i.e.,
$\operatorname*{dit}\left(  \pi\right)  -\operatorname*{dit}\left(
\sigma\right)  $, and then take the normalized counting measure of that subset
of $\operatorname*{dit}\left(  \pi\right)  -\operatorname*{dit}\left(
\sigma\right)  \subseteq U\times U$:

\begin{center}
$h\left(  \pi|\sigma\right)  =\frac{\left\vert \operatorname*{dit}\left(
\pi\right)  -\operatorname*{dit}\left(  \sigma\right)  \right\vert
}{\left\vert U\right\vert ^{2}}$

\textit{Logical conditional entropy of }$\pi$\textit{\ given }$\sigma$.
\end{center}

When the two partitions $\pi$ and $\sigma$ are joined together in the join
$\pi\vee\sigma$, whose blocks are the non-empty intersections $B\cap C$, their
information as distinctions is also joined together as sets,
$\operatorname*{dit}\left(  \pi\vee\sigma\right)  =\operatorname*{dit}\left(
\pi\right)  \cup\operatorname*{dit}\left(  \sigma\right)  $ (the "union"
mentioned by Campbell), which has the normalized counting measure of:

\begin{center}
$h\left(  \pi\vee\sigma\right)  =\frac{\left\vert \operatorname*{dit}\left(
\pi\right)  \cup\operatorname*{dit}\left(  \sigma\right)  \right\vert
}{\left\vert U\right\vert ^{2}}=\sum_{B\in\pi,C\in\sigma}p_{B\cap C}\left[
1-p_{B\cap C}\right]  $

\textit{logical entropy of a partition join} $\pi\vee\sigma$.
\end{center}

\noindent This logical entropy is interpreted as the probability that a pair
of random draws from $U$ will yield a $\pi$-distinction \textit{or} a $\sigma
$-distinction (where "or" includes both).

Then the relationships between the logical entropy concepts can be read off
the Venn diagram inclusion-exclusion principle for the dit sets:

\begin{center}
$\left\vert \operatorname*{dit}\left(  \pi\right)  \right\vert +\left\vert
\operatorname*{dit}\left(  \sigma\right)  \right\vert =\left\vert
\operatorname*{dit}\left(  \pi\vee\sigma\right)  \right\vert +\left\vert
\operatorname*{dit}\left(  \pi\right)  \cap\operatorname*{dit}\left(
\sigma\right)  \right\vert $
\end{center}

\noindent so that

\begin{center}
$\left\vert \operatorname*{dit}\left(  \pi\right)  -\operatorname*{dit}\left(
\sigma\right)  \right\vert =\left\vert \operatorname*{dit}\left(  \pi\right)
\right\vert -\left\vert \operatorname*{dit}\left(  \pi\right)  \cap
\operatorname*{dit}\left(  \sigma\right)  \right\vert =\left\vert
\operatorname*{dit}\left(  \pi\vee\sigma\right)  \right\vert -\left\vert
\operatorname*{dit}\left(  \sigma\right)  \right\vert $.%

\begin{center}
\includegraphics[
height=1.6042in,
width=2.4043in
]%
{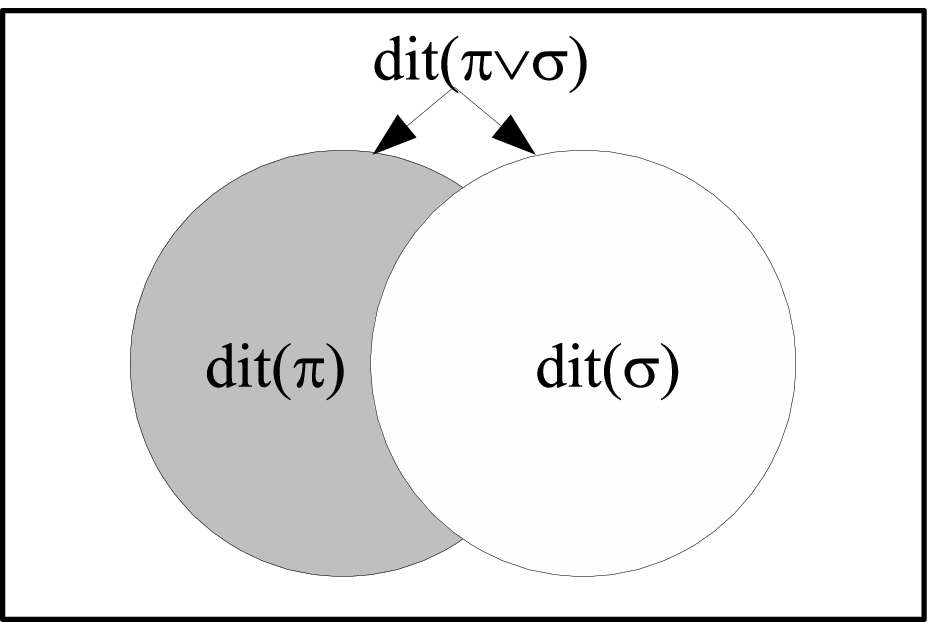}%
\end{center}

Figure 1: Venn diagram for subsets of $U\times U$
\end{center}

\noindent The shaded area in the Venn diagram has the dit-count measure:

\begin{center}
$\left\vert \operatorname*{dit}\left(  \pi\right)  -\operatorname*{dit}\left(
\sigma\right)  \right\vert =\left\vert \left(  \operatorname*{dit}\left(
\pi\right)  \cup\operatorname*{dit}\left(  \sigma\right)  \right)  \right\vert
-\left\vert \operatorname*{dit}\left(  \sigma\right)  \right\vert $

$h(\pi|\sigma)=h\left(  \pi\vee\sigma\right)  -h\left(  \sigma\right)  $.
\end{center}

For the corresponding definitions for random variables and their probability
distributions, consider a random variable $\left(  x,y\right)  $ taking values
in the product $X\times Y$ of finite sets with the joint probability
distribution $p\left(  x,y\right)  $, and thus with the marginal
distributions: $p\left(  x\right)  =\sum_{y\in Y}p\left(  x,y\right)  $ and
$p\left(  y\right)  =\sum_{x\in X}p\left(  x,y\right)  $. For notational
simplicity, the entropies can be considered as functions of the random
variables or of their probability distributions, e.g., $h\left(  p\left(
x,y\right)  \right)  =h\left(  x,y\right)  $. For the joint distribution, we
have the:

\begin{center}
$h\left(  x,y\right)  =h\left(  p\left(  x,y\right)  \right)  =\sum_{x\in
X,y\in Y}p\left(  x,y\right)  \left[  1-p\left(  x,y\right)  \right]  $

\textit{logical entropy of the joint distribution}
\end{center}

\noindent which is the probability that two samplings of the joint
distribution will yield a pair of \textit{distinct} ordered pairs $\left(
x,y\right)  $, $\left(  x^{\prime},y^{\prime}\right)  \in X\times Y$, i.e.,
with an $X$-distinction $x\neq x^{\prime}$ \textit{or} a $Y$-distinction
$y\neq y^{\prime}$.

For the definition of the conditional entropy $h\left(  x|y\right)  $, we
simply take the product measure of the set of pairs $\left(  x,y\right)  $ and
$\left(  x^{\prime},y^{\prime}\right)  $ that give an $X$-distinction but not
a $Y$-distinction. Thus given the first draw $\left(  x,y\right)  $, we can
again use a Venn diagram to compute the probability that the second draw
$\left(  x^{\prime},y^{\prime}\right)  $ will have $x^{\prime}\neq x$ but
$y^{\prime}=y$.

To illustrate this using Venn diagram reasoning, consider the probability
measure defined by $p\left(  x,y\right)  $ on the subsets of $X\times Y$.
Given the first draw $\left(  x,y\right)  $, the probability of getting an
$\left(  x,y\right)  $-distinction on the second draw is $1-p\left(
x,y\right)  $ and the probability of getting a $y$-distinction is $1-p\left(
y\right)  $. A draw that is a $y$-distinction is, a fortiori, an $\left(
x,y\right)  $-distinction so the area $1-p\left(  y\right)  $ is contained in
the area $1-p\left(  x,y\right)  $. Then the probability of getting an
$\left(  x,y\right)  $-distinction that is not a $y$-distinction on the second
draw is the difference: $\left(  1-p\left(  x,y\right)  \right)  -\left(
1-p\left(  y\right)  \right)  =p\left(  y\right)  -p\left(  x,y\right)  $.%

\begin{center}
\includegraphics[
height=1.7669in,
width=2.3271in
]%
{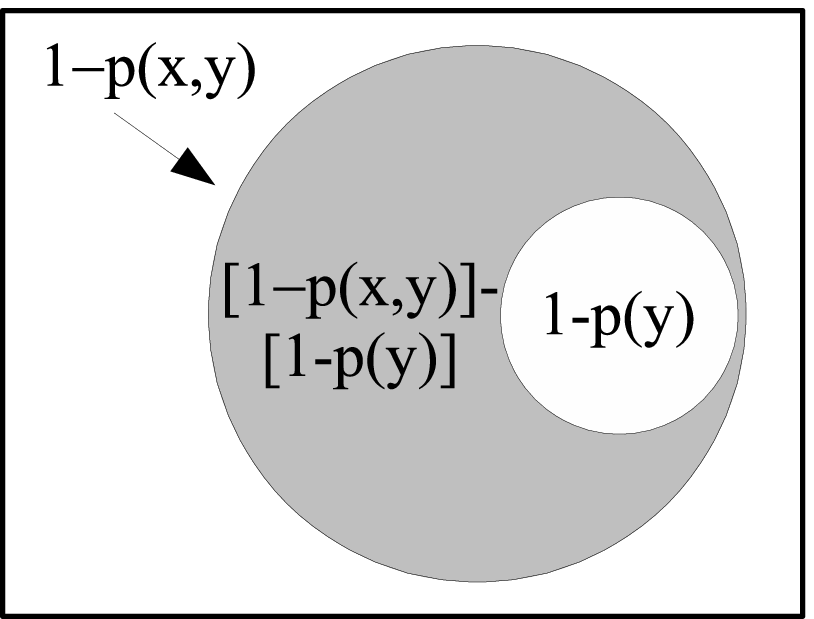}%
\end{center}

\begin{center}
Figure 2: $\left(  1-p\left(  x,y\right)  \right)  -\left(  1-p\left(
y\right)  \right)  $

= probability of an $x$-distinction but not a $y$-distinction on $X\times Y$.
\end{center}

\noindent Since the first draw $\left(  x,y\right)  $ was with probability
$p\left(  x,y\right)  $, we have the following as the product measure of the
subset of $\left[  X\times Y\right]  ^{2}$ of pairs $\left[  \left(
x,y\right)  ,\left(  x^{\prime},y^{\prime}\right)  \right]  $ that are
$X$-distinctions but not $Y$-distinctions:

\begin{center}
$h\left(  x|y\right)  =\sum_{x,y}p\left(  x,y\right)  \left[  \left(
1-p\left(  x,y\right)  \right)  -\left(  1-p\left(  y\right)  \right)
\right]  $

\textit{logical conditional entropy of }$x$\textit{\ given }$y$.
\end{center}

\noindent Then a little algebra quickly yields:

\begin{center}
$h\left(  x|y\right)  =\sum_{x,y}p\left(  x,y\right)  \left[  \left(
1-p\left(  x,y\right)  \right)  -\left(  1-p\left(  y\right)  \right)
\right]  $

$=\left[  1-\sum_{x,y}p\left(  x,y\right)  ^{2}\right]  -\left[  1-\sum
_{y}p\left(  y\right)  ^{2}\right]  =h\left(  x,y\right)  -h\left(  y\right)
$.
\end{center}

The summation over $p\left(  x,y\right)  $ recasts the Venn diagram to the set
$\left(  X\times Y\right)  ^{2}$ where the product probability measure (for
the two independent draws) gives the logical entropies:

\begin{center}%
\begin{center}
\includegraphics[
height=1.8273in,
width=2.2642in
]%
{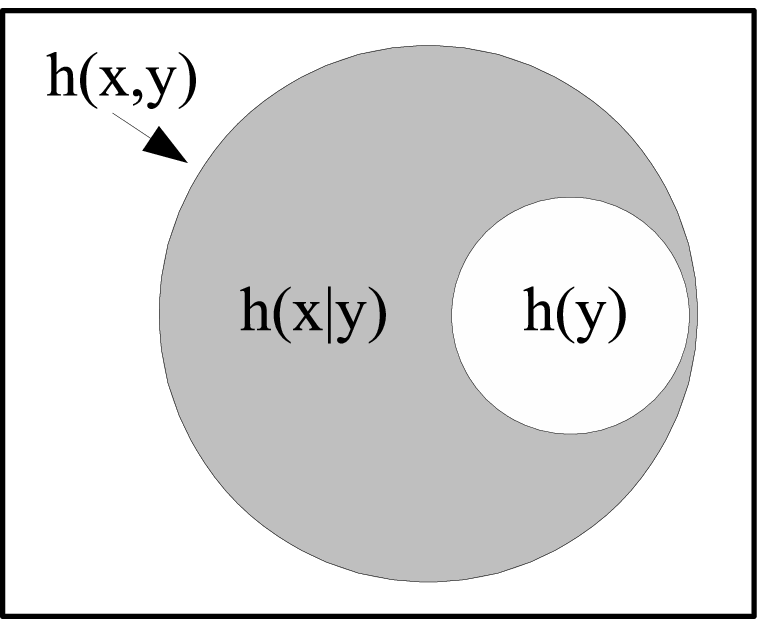}%
\end{center}

Figure 3: $h\left(  x|y\right)  =h\left(  x,y\right)  -h\left(  y\right)  $.
\end{center}

It might be noted that the logical conditional entropy, like the other logical
entropies, is not just an average; the conditional entropy is the product
probability measure of the subset:

\begin{center}
$\left\{  \left[  \left(  x,y\right)  ,\left(  x^{\prime},y^{\prime}\right)
\right]  :x\neq x^{\prime},y=y^{\prime}\right\}  \subseteq\left(  X\times
Y\right)  \times\left(  X\times Y\right)  $.
\end{center}

\subsection{Shannon conditional entropy}

The Shannon conditional entropy for partitions $\pi$ and $\sigma$ is based on
subset reasoning which is then averaged over a partition. Given a subset $C$
$\in\sigma$, a partition $\pi=\left\{  B\right\}  _{B\in\pi}$ induces a
partition of $C$ with the blocks $\left\{  B\cap C\right\}  _{B\in\pi}$. Then
$p_{B|C}=\frac{p_{B\cap C}}{p_{C}}$ is the probability distribution associated
with that partition so it has a Shannon entropy which we denote: $H\left(
\pi|C\right)  =\sum_{B\in\pi}p_{B|C}\log\left(  \frac{1}{p_{B|C}}\right)
=\sum_{B}\frac{p_{B\cap C}}{p_{C}}\log\left(  \frac{p_{C}}{p_{B\cap C}%
}\right)  $. The Shannon conditional entropy is then obtained by averaging
over the blocks of $\sigma$:

\begin{center}
$H\left(  \pi|\sigma\right)  =\sum_{C\in\sigma}p_{C}H\left(  \pi|C\right)
=\sum_{B,C}p_{B\cap C}\log\left(  \frac{p_{C}}{p_{B\cap C}}\right)  $

\textit{Shannon conditional entropy of }$\pi$\textit{\ given }$\sigma$.
\end{center}

Since the join $\pi\vee\sigma$ is the partition whose blocks are the non-empty
intersections $B\cap C$,

\begin{center}
$H\left(  \pi\vee\sigma\right)  =\sum_{B,C}p_{B\cap C}\log\left(  \frac
{1}{p_{B\cap C}}\right)  $.
\end{center}

\noindent Developing the formula gives:

\begin{center}
$H\left(  \pi|\sigma\right)  =\sum_{C}\left[  p_{C}\log\left(  p_{C}\right)
-\sum_{B}p_{B\cap C}\log\left(  p_{B\cap C}\right)  \right]  =H\left(  \pi
\vee\sigma\right)  -H\left(  \sigma\right)  $.
\end{center}

\noindent Thus the conditional entropy $H\left(  \pi|\sigma\right)  $ is
interpreted as the Shannon-information contained in the join $\pi\vee\sigma$
that is not contained in $\sigma$.%

\begin{center}
\includegraphics[
height=1.7435in,
width=2.1821in
]%
{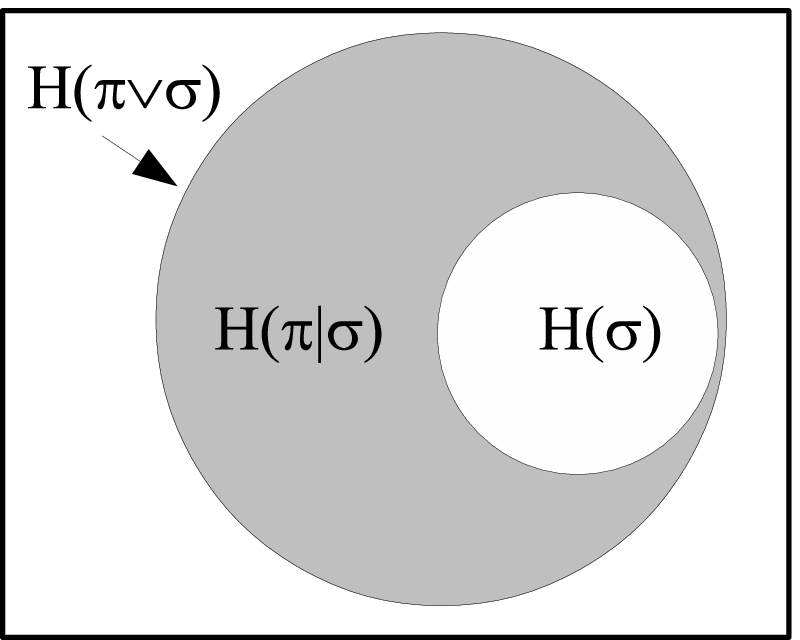}%
\end{center}

\begin{center}
Figure 4: $H\left(  \pi|\sigma\right)  =H\left(  \pi\vee\sigma\right)
-H\left(  \sigma\right)  $

"Venn diagram picture" for Shannon conditional entropy of partitions
\end{center}

Given the joint distribution $p\left(  x,y\right)  $ on $X\times Y$, the
conditional probability distribution for a specific $y_{0}\in Y$ is $p\left(
x|y_{0}\right)  =\frac{p\left(  x,y_{0}\right)  }{p\left(  y_{0}\right)  }$
which has the Shannon entropy: $H\left(  x|y_{0}\right)  =\sum_{x}p\left(
x|y_{0}\right)  \log\left(  \frac{1}{p\left(  x|y_{0}\right)  }\right)  $.
Then the conditional entropy is the average of these entropies:

\begin{center}
$H\left(  x|y\right)  =\sum_{y}p\left(  y\right)  \sum_{x}\frac{p\left(
x,y\right)  }{p\left(  y\right)  }\log\left(  \frac{p\left(  y\right)
}{p\left(  x,y\right)  }\right)  =\sum_{x,y}p\left(  x,y\right)  \log\left(
\frac{p\left(  y\right)  }{p\left(  x,y\right)  }\right)  $

\textit{Shannon conditional entropy of }$x$\textit{\ given }$y$.
\end{center}

\noindent Expanding as before gives $H\left(  x|y\right)  =H\left(
x,y\right)  -H\left(  y\right)  $ with a similar Venn diagram picture (see below).

\subsection{Shannon conditional entropy from logical conditional entropy}

Now we can develop the Shannon conditional entropy from the logical
conditional entropy and thereby explain the Venn diagram relationship. The
logical conditional entropy is:

\begin{center}
$h\left(  x|y\right)  =\sum_{x,y}p\left(  x,y\right)  \left[  \left(
1-p\left(  x,y\right)  \right)  -\left(  1-p\left(  y\right)  \right)
\right]  $
\end{center}

\noindent where $1-p\left(  x,y\right)  $ is the normalized dit count for the
discrete partition on a set $U_{1/p\left(  x,y\right)  }$ with $\frac
{1}{p\left(  x,y\right)  }$ equiprobable elements. Hence that same
equiprobable set requires the bit-count of $\log\left(  \frac{1}{p\left(
x,y\right)  }\right)  $ binary partitions to distinguish its elements.
Similarly $1-p\left(  y\right)  $ is the normalized dit count for (the
discrete partition on) a set $U_{1/p\left(  y\right)  }$ with $\frac
{1}{p\left(  y\right)  }$ equiprobable elements, so it requires $\log\left(
\frac{1}{p\left(  y\right)  }\right)  $ binary partitions to make those
distinctions. Those binary partitions are included in the $\log\left(
\frac{1}{p\left(  x,y\right)  }\right)  $ binary partitions (since a
$y$-distinction is automatically a $\left(  x,y\right)  $-distinction) and we
don't want the $y$-distinctions so they are subtracted off to get:
$\log\left(  \frac{1}{p\left(  x,y\right)  }\right)  -\log\left(  \frac
{1}{p\left(  y\right)  }\right)  $ bits. Taking the same probabilistic
average, the average number of binary partitions needed to make the
$x$-distinctions but not the $y$-distinctions is:

\begin{center}
$\sum_{x,y}p\left(  x,y\right)  \left[  \log\left(  \frac{1}{p\left(
x,y\right)  }\right)  -\log\left(  \frac{1}{p\left(  y\right)  }\right)
\right]  =\sum_{x,y}p\left(  x,y\right)  \log\left(  \frac{p\left(  y\right)
}{p\left(  x,y\right)  }\right)  =H\left(  x|y\right)  .$
\end{center}

\noindent Replacing the dit-counts by the bit-counts for the equiprobable
sets, and taking the probabilistic average gives the same Venn diagram picture
for the Shannon entropies.

\begin{center}%
\begin{center}
\includegraphics[
height=1.8802in,
width=2.3724in
]%
{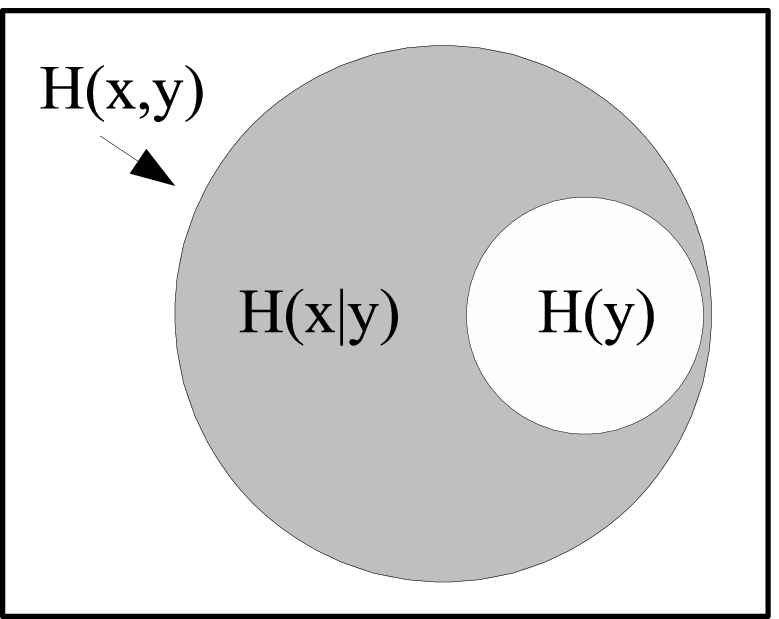}%
\end{center}

Figure 5: $H\left(  x|y\right)  =H\left(  x,y\right)  -H\left(  y\right)  $.
\end{center}

\section{Mutual information for logical entropies}

\subsection{The case for partitions}

If the \textquotedblleft atom\textquotedblright\ of information is the
distinction or dit, then the atomic information in a partition $\pi$ is\ its
dit set, $\operatorname*{dit}(\pi)$. Following again Campbell's dictum about
the mutual information, the information common to two partitions $\pi$ and
$\sigma$ would naturally be the intersection of their$\ $dit sets:

\begin{center}
$\operatorname*{Mut}(\pi,\sigma)=\operatorname*{dit}\left(  \pi\right)
\cap\operatorname*{dit}\left(  \sigma\right)  $

\textit{Mutual information set}.
\end{center}

\noindent It is an interesting and not completely trivial fact that as long as
neither $\pi$ nor $\sigma$ are the indiscrete partition $\mathbf{0}$ (where
$\operatorname*{dit}\left(  \mathbf{0}\right)  =\emptyset$), then $\pi$ and
$\sigma$ have a distinction in common.

\begin{proposition}
[Non-empty dit sets intersect]Given two partitions $\pi$ and $\sigma$ on $U$
with non-empty dit sets, $\operatorname*{dit}\left(  \pi\right)
\cap\operatorname*{dit}\left(  \sigma\right)  \neq\emptyset$.\footnote{The
contrapositive of the "non-empty dit sets intersect" proposition is also
interesting. Given two equivalence relations $E_{1},E_{2}\subseteq U^{2}$, if
every pair of elements $u,u^{\prime}\in U$ is equated by one or the other of
the relations, i.e., $E_{1}\cup E_{2}=U^{2}$, then either $E_{1}=U^{2}$ or
$E_{2}=U^{2}$.}
\end{proposition}

\noindent Since $\pi$ is not the indiscrete partition, consider two elements
$u$ and $u^{\prime}$ distinguished by $\pi$ but identified by $\sigma$
[otherwise $\left(  u,u^{\prime}\right)  \in\operatorname*{dit}\left(
\pi\right)  \cap\operatorname*{dit}\left(  \sigma\right)  $]. Since $\sigma$
is also not the indiscrete partition, there must be a third element
$u^{\prime\prime}$ not in the same block of $\sigma$ as $u$ and $u^{\prime}$.
But since $u$ and $u^{\prime}$ are in different blocks of $\pi$, the third
element $u^{\prime\prime}$ must be distinguished from one or the other or both
in $\pi$. Hence $\left(  u,u^{\prime\prime}\right)  $ or $\left(  u^{\prime
},u^{\prime\prime}\right)  $ must be distinguished by both partitions and thus
must be in their mutual information set $\operatorname*{Mut}\left(  \pi
,\sigma\right)  =\operatorname*{dit}\left(  \pi\right)  \cap
\operatorname*{dit}\left(  \sigma\right)  $.$\square$

The dit sets $\operatorname*{dit}\left(  \pi\right)  $ and their complementary
indit sets (= equivalence relations) $\operatorname*{indit}\left(  \pi\right)
=U^{2}-\operatorname*{dit}\left(  \pi\right)  $ are easily characterized as:%

\begin{align*}
\operatorname*{indit}\left(  \pi\right)   &  =%
{\textstyle\bigcup\limits_{B\in\pi}}
B\times B\\
\operatorname*{dit}\left(  \pi\right)   &  =%
{\textstyle\bigcup\limits_{B\neq B^{\prime};B,B^{\prime}\in\pi}}
B\times B^{\prime}=U\times U-\operatorname*{indit}\left(  \pi\right)
=\operatorname*{indit}\left(  \pi\right)  ^{c}.
\end{align*}

\noindent The mutual information set can also be characterized in this manner.

\begin{proposition}
[Structure of mutual information sets]Given partitions $\pi$ and $\sigma$ with
blocks $\left\{  B\right\}  _{B\in\pi}$ and $\left\{  C\right\}  _{C\in\sigma
}$, then
\end{proposition}

\begin{center}
$\operatorname*{Mut}\left(  \pi,\sigma\right)  =%
{\textstyle\bigcup\limits_{B\in\pi,C\in\sigma}}
\left(  B-\left(  B\cap C\right)  \right)  \times\left(  C-\left(  B\cap
C\right)  \right)  =\bigcup\limits_{B\in\pi,C\in\sigma}\left(  B-C\right)
\times\left(  C-B\right)  $.
\end{center}

\noindent The union (which is a disjoint union) will include the pairs
$\left(  u,u^{\prime}\right)  $ where for some $B\in\pi$ and $C\in\sigma$,
$u\in B-\left(  B\cap C\right)  $ and $u^{\prime}\in C-\left(  B\cap C\right)
$. Since $u^{\prime}$ is in $C$ but not in the intersection $B\cap C$, it must
be in a different block of $\pi$ than $B$ so $\left(  u,u^{\prime}\right)
\in\operatorname*{dit}\left(  \pi\right)  $. Symmetrically, $\left(
u,u^{\prime}\right)  \in\operatorname*{dit}\left(  \sigma\right)  $ so
$\left(  u,u^{\prime}\right)  \in\operatorname*{Mut}\left(  \pi,\sigma\right)
=\operatorname*{dit}\left(  \pi\right)  \cap\operatorname*{dit}\left(
\sigma\right)  $. Conversely if $\left(  u,u^{\prime}\right)  \in
\operatorname*{Mut}\left(  \pi,\sigma\right)  $ then take the $B$ containing
$u$ and the $C$ containing $u^{\prime}$. Since $\left(  u,u^{\prime}\right)  $
is distinguished by both partitions, $u\not \in C$ and $u^{\prime}\not \in B$
so that $\left(  u,u^{\prime}\right)  \in\left(  B-\left(  B\cap C\right)
\right)  \times\left(  C-\left(  B\cap C\right)  \right)  $.$\square$

The probability that a pair randomly chosen from $U\times U$ would be
distinguished by $\pi$ \textit{and} $\sigma$ would be given by the normalized
counting measure of the mutual information set which is the:

\begin{center}
$m(\pi,\sigma)=$ $\frac{\left\vert \operatorname*{dit}\left(  \pi\right)
\cap\operatorname*{dit}\left(  \sigma\right)  \right\vert }{\left\vert
U\right\vert ^{2}}=$ probability that $\pi$ and $\sigma$ distinguishes

\textit{Mutual logical information of }$\pi$\textit{\ and }$\sigma$.
\end{center}

By the inclusion-exclusion principle:

\begin{center}
$\left\vert \operatorname*{Mut}\left(  \pi,\sigma\right)  \right\vert
=\left\vert \operatorname*{dit}\left(  \pi\right)  \cap\operatorname*{dit}%
\left(  \sigma\right)  \right\vert =\left\vert \operatorname*{dit}\left(
\pi\right)  \right\vert +\left\vert \operatorname*{dit}\left(  \sigma\right)
\right\vert -\left\vert \operatorname*{dit}\left(  \pi\right)  \cup
\operatorname*{dit}\left(  \sigma\right)  \right\vert $.
\end{center}

\noindent Normalizing, the probability that a random pair is distinguished by
both partitions is given by the inclusion-exclusion principle:%

\begin{align*}
m\left(  \pi,\sigma\right)   &  =\frac{\left\vert \operatorname*{dit}\left(
\pi\right)  \cap\operatorname*{dit}\left(  \sigma\right)  \right\vert
}{\left\vert U\right\vert ^{2}}\\
&  =\frac{\left\vert \operatorname*{dit}\left(  \pi\right)  \right\vert
}{\left\vert U\right\vert ^{2}}+\frac{\left\vert \operatorname*{dit}\left(
\sigma\right)  \right\vert }{\left\vert U\right\vert ^{2}}-\frac{\left\vert
\operatorname*{dit}\left(  \pi\right)  \cup\operatorname*{dit}\left(
\sigma\right)  \right\vert }{\left\vert U\right\vert ^{2}}\\
&  =h\left(  \pi\right)  +h\left(  \sigma\right)  -h\left(  \pi\vee
\sigma\right)  \text{.}%
\end{align*}

\begin{center}
Inclusion-exclusion principle for logical entropies of partitions
\end{center}

\noindent This can be extended after the fashion of the inclusion-exclusion
principle to any number of partitions. It was previously noted that the
intersection of two dit sets is not necessarily the dit set of a partition,
but the interior of the intersection is the dit set $\operatorname*{dit}%
\left(  \pi\wedge\sigma\right)  $ of the partition meet $\pi\wedge\sigma$.
Hence we also have the:

\begin{center}
$h\left(  \pi\wedge\sigma\right)  \leq h\left(  \pi\right)  +h\left(
\sigma\right)  -h\left(  \pi\vee\sigma\right)  $

Submodular inequality for logical entropies.
\end{center}

\subsection{The case for joint distributions}

Consider again a joint distribution $p\left(  x,y\right)  $ over $X\times Y$
for finite $X$ and $Y$. Intuitively, the mutual logical information $m\left(
x,y\right)  $ in the joint distribution $p\left(  x,y\right)  $ would be the
probability that a sampled pair $\left(  x,y\right)  $ would be a distinction
of $p\left(  x\right)  $ \textit{and }a distinction of $p\left(  y\right)  $.
That means for each probability $p\left(  x,y\right)  $, it must be multiplied
by the probability of not drawing the same $x$ \textit{and} not drawing the
same $y$ (e.g., in a second independent drawing). In the Venn diagram, the
area or probability of the drawing that $x$ or that $y$ is $p\left(  x\right)
+p\left(  y\right)  -p\left(  x,y\right)  $ (correcting for adding the overlap
twice) so the probability of getting neither that $x$ nor that $y$ is the
complement $1-p\left(  x\right)  -p\left(  y\right)  +p\left(  x,y\right)
=\left[  1-p\left(  x\right)  \right]  +\left[  1-p\left(  y\right)  \right]
-\left[  1-p\left(  x,y\right)  \right]  $.%

\begin{center}
\includegraphics[
height=1.5489in,
width=2.3431in
]%
{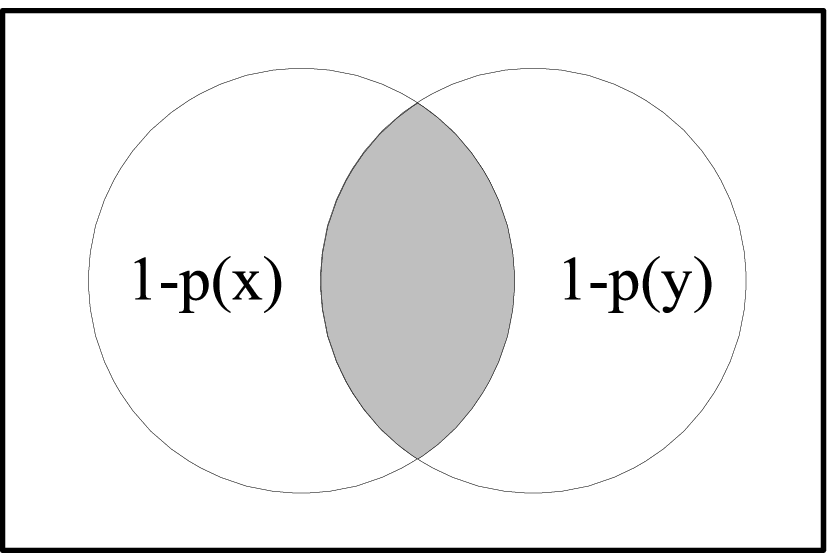}%
\end{center}

\begin{center}
Figure 6: $\left[  1-p\left(  x\right)  \right]  +\left[  1-p\left(  y\right)
\right]  -\left[  1-p\left(  x,y\right)  \right]  $

= shaded area in Venn diagram for $X\times Y$
\end{center}

Hence we have:

\begin{center}
$m\left(  x,y\right)  =\sum_{x,y}p\left(  x,y\right)  \left[  \left[
1-p\left(  x\right)  \right]  +\left[  1-p\left(  y\right)  \right]  -\left[
1-p\left(  x,y\right)  \right]  \right]  $

\textit{Logical mutual information in a joint probability distribution}.
\end{center}

The probability of two independent draws differing in \textit{either} the $x$
\textit{or} the $y$ is just the logical entropy of the joint distribution:

\begin{center}
$h\left(  x,y\right)  =h\left(  p\left(  x,y\right)  \right)  =\sum
_{x,y}p\left(  x,y\right)  \left[  1-p\left(  x,y\right)  \right]
=1-\sum_{x,y}p\left(  x,y\right)  ^{2}$.
\end{center}

\noindent Using a little algebra to expand the logical mutual information:%

\begin{align*}
m\left(  x,y\right)   &  =\left[  1-%
{\textstyle\sum\nolimits_{x,y}}
p\left(  x,y\right)  p\left(  x\right)  \right]  +\left[  1-%
{\textstyle\sum\nolimits_{x,y}}
p\left(  x,y\right)  p\left(  y\right)  \right]  -\left[  1-%
{\textstyle\sum\nolimits_{x,y}}
p\left(  x,y\right)  ^{2}\right] \\
&  =h\left(  x\right)  +h\left(  y\right)  -h\left(  x,y\right)
\end{align*}

\begin{center}
Inclusion-exclusion principle for logical entropies of a joint distribution.%

\begin{center}
\includegraphics[
height=1.8642in,
width=2.5326in
]%
{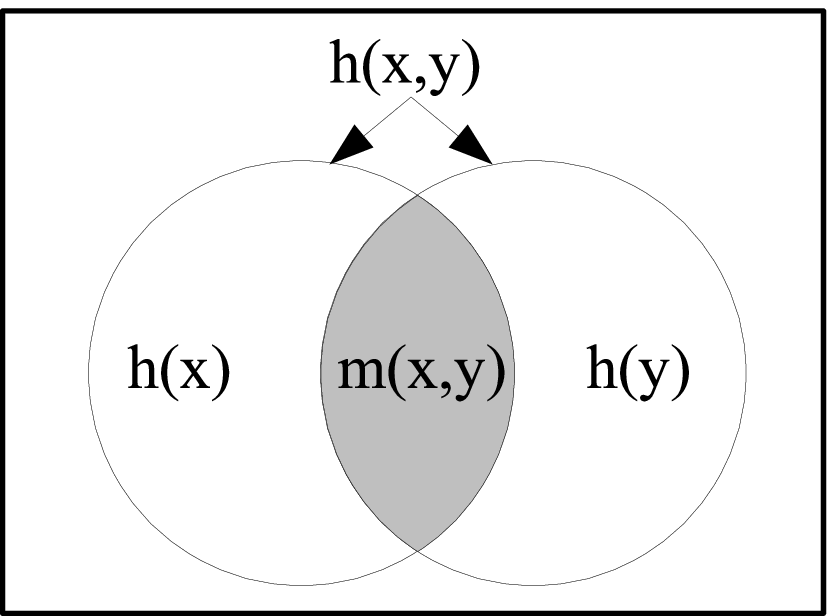}%
\end{center}

Figure 7: $m\left(  x,y\right)  =h\left(  x\right)  +h\left(  y\right)
-h\left(  x,y\right)  $

= shaded area in Venn diagram for $\left(  X\times Y\right)  ^{2}$.
\end{center}

It might be noted that the logical mutual information, like the other logical
entropies, is not just an average; the mutual information is the product
probability measure of the subset:

\begin{center}
$\left\{  \left[  \left(  x,y\right)  ,\left(  x^{\prime},y^{\prime}\right)
\right]  :x\neq x^{\prime},y\neq y^{\prime}\right\}  \subseteq\left(  X\times
Y\right)  \times\left(  X\times Y\right)  $.
\end{center}

\section{Mutual information for Shannon entropies}

\subsection{The case for partitions}

The usual heuristic motivation for Shannon's mutual information is much like
its development from the logical mutual information so we will take that
approach at the outset. The logical mutual information for partitions can be
expressed in the form:

\begin{center}
$m\left(  \pi,\sigma\right)  =\sum_{B,C}p_{B\cap C}\left[  \left(
1-p_{B}\right)  +\left(  1-p_{C}\right)  -\left(  1-p_{B\cap C}\right)
\right]  $
\end{center}

\noindent so if we substitute the bit-counts for the dit-counts as before, we get:

\begin{center}
$I\left(  \pi,\sigma\right)  =\sum_{B,C}p_{B\cap C}\left[  \log\left(
\frac{1}{p_{B}}\right)  +\log\left(  \frac{1}{p_{C}}\right)  -\log\left(
\frac{1}{p_{B\cap C}}\right)  \right]  =\sum_{B,C}p_{B\cap C}\log\left(
\frac{p_{B\cap C}}{p_{B}p_{C}}\right)  $

\textit{Shannon's mutual information for partitions}.
\end{center}

Keeping the log's separate gives the Venn diagram picture:%

\begin{align*}
I\left(  \pi,\sigma\right)   &  =\sum_{B,C}p_{B\cap C}\left[  \log\left(
\frac{1}{p_{B}}\right)  +\log\left(  \frac{1}{p_{C}}\right)  -\log\left(
\frac{1}{p_{B\cap C}}\right)  \right] \\
&  =H\left(  \pi\right)  +H\left(  \sigma\right)  -H\left(  \pi\vee
\sigma\right)
\end{align*}

\begin{center}
Inclusion-exclusion analogy for Shannon entropies of partitions.
\end{center}

\subsection{The case for joint distributions}

To move from partitions to probability distributions, consider again the joint
distribution $p\left(  x,y\right)  $ on $X\times Y$. Then developing the
Shannon mutual information from the logical mutual information amounts to
replacing the block probabilities $p_{B\cap C}$ in the join $\pi\vee\sigma$ by
the joint probabilities $p\left(  x,y\right)  $ and the probabilities in the
separate partitions by the marginals (since $p_{B}=\sum_{C\in\sigma}p_{B\cap
C}$ and $p_{C}=\sum_{B\in\pi}p_{B\cap C}$), to obtain:

\begin{center}
$I\left(  x,y\right)  =\sum_{x,y}p\left(  x,y\right)  \log\left(
\frac{p\left(  x,y\right)  }{p\left(  x\right)  p\left(  y\right)  }\right)  $

\textit{Shannon mutual information in a joint probability distribution}.
\end{center}

Then the same proof carries over to give the:

\begin{center}
$I\left(  x,y\right)  =H\left(  x\right)  +H\left(  y\right)  -H\left(
x,y\right)  $%

\begin{center}
\includegraphics[
height=1.7309in,
width=2.3506in
]%
{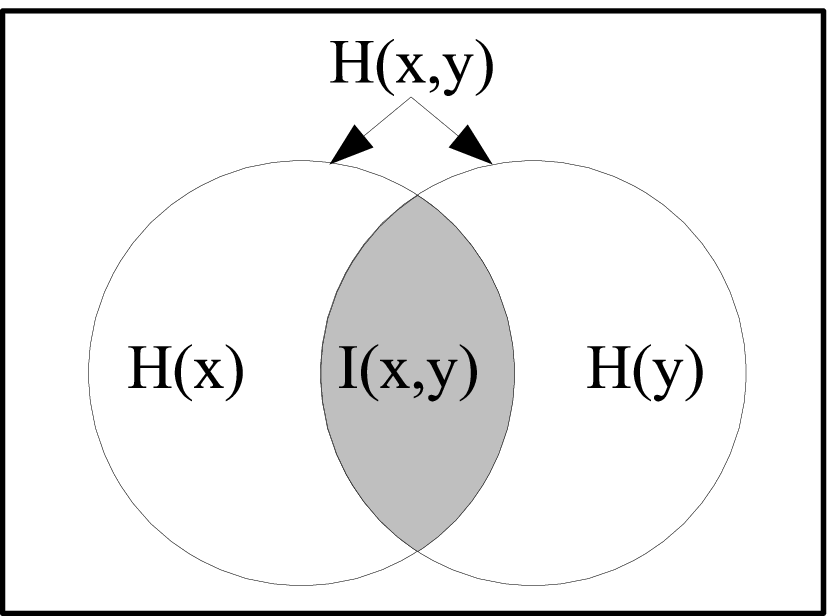}%
\end{center}

Figure 8: Inclusion-exclusion "picture" for Shannon entropies of probability distributions.
\end{center}

The logical mutual information formula:

\begin{center}
$m\left(  x,y\right)  =\sum_{x,y}p\left(  x,y\right)  \left[  \left[
1-p\left(  x\right)  \right]  +\left[  1-p\left(  y\right)  \right]  -\left[
1-p\left(  x,y\right)  \right]  \right]  $
\end{center}

\noindent develops via the dit-count to bit-count conversion to:

\begin{center}
$\sum_{x,y}p\left(  x,y\right)  \left[  \log\left(  \frac{1}{p\left(
x\right)  }\right)  +\log\left(  \frac{1}{p\left(  y\right)  }\right)
-\log\left(  \frac{1}{p\left(  x,y\right)  }\right)  \right]  =\sum
_{x,y}p\left(  x,y\right)  \log\left(  \frac{p\left(  x,y\right)  }{p\left(
x\right)  p\left(  y\right)  }\right)  =I\left(  x,y\right)  $.
\end{center}

Thus the genuine Venn diagram relationships for the product probability
measure that gives the logical entropies carry over, via the dit-count to
bit-count conversion, to give a similar Venn diagram picture for the Shannon entropies.

\section{Independence}

\subsection{Independent Partitions}

Two partitions $\pi$ and $\sigma$ are said to be (stochastically)
\textit{independent} if for all $B\in\pi$ and $C\in\sigma$, $p_{B\cap C}%
=p_{B}p_{C}$. If $\pi$ and $\sigma$ are independent, then:

\begin{center}
$I\left(  \pi,\sigma\right)  =\sum_{B\in\pi,C\in\sigma}p_{B\cap C}\log\left(
\frac{p_{B\cap C}}{p_{B}p_{C}}\right)  =0=H\left(  \pi\right)  +H\left(
\sigma\right)  -H\left(  \pi\vee\sigma\right)  $,
\end{center}

\noindent so that:

\begin{center}
$H\left(  \pi\vee\sigma\right)  =H\left(  \pi\right)  +H\left(  \sigma\right)
$

Shannon entropy for partitions additive under independence.
\end{center}

In ordinary probability theory, two events $E,E^{\prime}\subseteq U$ for a
sample space $U$ are said to be \textit{independent} if $\Pr\left(  E\cap
E^{\prime}\right)  =\Pr\left(  E\right)  \Pr\left(  E^{\prime}\right)  $. We
have used the motivation of thinking of a partition-as-dit-set
$\operatorname*{dit}\left(  \pi\right)  $ as an \textquotedblleft
event\textquotedblright\ in a sample space $U\times U$ with the probability of
that event being $h\left(  \pi\right)  $, the logical entropy of the
partition. The following proposition shows that this motivation extends to the
notion of independence.

\begin{proposition}
[Independent partitions have independent dit sets]If $\pi$ and $\sigma$ are
(stochastically) independent partitions, then their dit sets
$\operatorname*{dit}\left(  \pi\right)  $ and $\operatorname*{dit}\left(
\sigma\right)  $ are independent as events in the sample space $U\times U$
(with equiprobable points).
\end{proposition}

\noindent For independent partitions $\pi$ and $\sigma$, we need to show that
the probability $m(\pi,\sigma)$ of the event $\operatorname*{Mut}\left(
\pi,\sigma\right)  =\operatorname*{dit}\left(  \pi\right)  \cap
\operatorname*{dit}\left(  \sigma\right)  $ is equal to the product of the
probabilities $h\left(  \pi\right)  $ and $h\left(  \sigma\right)  $ of the
events $\operatorname*{dit}\left(  \pi\right)  $ and $\operatorname*{dit}%
\left(  \sigma\right)  $ in the sample space $U\times U$. By the assumption of
stochastic independence, we have $\frac{\left\vert B\cap C\right\vert
}{\left\vert U\right\vert }=p_{B\cap C}=p_{B}p_{C}=\frac{\left\vert
B\right\vert \left\vert C\right\vert }{\left\vert U\right\vert ^{2}}$ so that
$\left\vert B\cap C\right\vert =\left\vert B\right\vert \left\vert
C\right\vert /\left\vert U\right\vert $. By the previous structure theorem for
the mutual information set: $\operatorname*{Mut}\left(  \pi,\sigma\right)  =%
{\textstyle\bigcup\limits_{B\in\pi,C\in\sigma}}
\left(  B-\left(  B\cap C\right)  \right)  \times\left(  C-\left(  B\cap
C\right)  \right)  $, where the union is disjoint so that:

\begin{center}%
\begin{align*}
\left\vert \operatorname*{Mut}\left(  \pi,\sigma\right)  \right\vert  &  =%
{\textstyle\sum\nolimits_{B\in\pi,C\in\sigma}}
\left(  \left\vert B\right\vert -\left\vert B\cap C\right\vert \right)
\left(  \left\vert C\right\vert -\left\vert B\cap C\right\vert \right) \\
&  =%
{\textstyle\sum\nolimits_{B\in\pi,C\in\sigma}}
\left(  \left\vert B\right\vert -\frac{|B|\left\vert C\right\vert }{\left\vert
U\right\vert }\right)  \left(  \left\vert C\right\vert -\frac{|B|\left\vert
C\right\vert }{\left\vert U\right\vert }\right) \\
&  =\frac{1}{\left\vert U\right\vert ^{2}}%
{\textstyle\sum\nolimits_{B\in\pi,C\in\sigma}}
\left\vert B\right\vert \left(  \left\vert U\right\vert -\left\vert
C\right\vert \right)  \left\vert C\right\vert \left(  \left\vert U\right\vert
-\left\vert B\right\vert \right) \\
&  =\frac{1}{\left\vert U\right\vert ^{2}}%
{\textstyle\sum\nolimits_{B\in\pi}}
\left\vert B\right\vert \left\vert U-B\right\vert
{\textstyle\sum\nolimits_{C\in\sigma}}
\left\vert C\right\vert \left\vert U-C\right\vert \\
&  =\frac{1}{\left\vert U\right\vert ^{2}}\left\vert \operatorname*{dit}%
\left(  \pi\right)  \right\vert \left\vert \operatorname*{dit}\left(
\sigma\right)  \right\vert
\end{align*}

\end{center}

\noindent so that:

\begin{center}
$m(\pi,\sigma)=\frac{\left\vert \operatorname*{Mut}\left(  \pi,\sigma\right)
\right\vert }{\left\vert U\right\vert ^{2}}=\frac{\left\vert
\operatorname*{dit}\left(  \pi\right)  \right\vert }{\left\vert U\right\vert
^{2}}\frac{\left\vert \operatorname*{dit}\left(  \sigma\right)  \right\vert
}{\left\vert U\right\vert ^{2}}=h\left(  \pi\right)  h\left(  \sigma\right)
$.$\square$
\end{center}

\noindent Hence the logical entropies behave like probabilities under
independence; the probability that $\pi$ \textit{and} $\sigma$ distinguishes,
i.e., $m\left(  \pi,\sigma\right)  $, is equal to the probability $h\left(
\pi\right)  $ that $\pi$ distinguishes times the probability $h\left(
\sigma\right)  $ that $\sigma$ distinguishes:

\begin{center}
$m(\pi,\sigma)=h\left(  \pi\right)  h\left(  \sigma\right)  $

Logical entropy multiplicative under independence.
\end{center}

It is sometimes convenient to think in the complementary terms of an
equivalence relation "equating" or \textquotedblleft
identifying\textquotedblright\ rather than a partition distinguishing. Since
$h\left(  \pi\right)  $ can be interpreted as the probability that a random
pair of elements from $U$ are distinguished by $\pi$, i.e., as a distinction
probability, its complement $1-h\left(  \pi\right)  $ can be interpreted as an
\textit{identification probability}, i.e., the probability that a random pair
is equated by $\pi$ (thinking of $\pi$ as an equivalence relation on $U$). In general,

\begin{center}
$\left[  1-h\left(  \pi\right)  \right]  \left[  1-h\left(  \sigma\right)
\right]  =1-h\left(  \pi\right)  -h\left(  \sigma\right)  +h\left(
\pi\right)  h\left(  \sigma\right)  =\left[  1-h\left(  \pi\vee\sigma\right)
\right]  +\left[  h\left(  \pi\right)  h\left(  \sigma\right)  -m(\pi
,\sigma\right]  $
\end{center}

\noindent which could also be rewritten as:

\begin{center}
$\left[  1-h\left(  \pi\vee\sigma\right)  \right]  -\left[  1-h\left(
\pi\right)  \right]  \left[  1-h\left(  \sigma\right)  \right]  =m(\pi
,\sigma)-h\left(  \pi\right)  h\left(  \sigma\right)  $.
\end{center}

\noindent Thus if $\pi$ and $\sigma$ are independent, then the probability
that the join partition $\pi\vee\sigma$ identifies is the probability that
$\pi$ identifies times the probability that $\sigma$ identifies:

\begin{center}
$\left[  1-h\left(  \pi\right)  \right]  \left[  1-h\left(  \sigma\right)
\right]  =\left[  1-h\left(  \pi\vee\sigma\right)  \right]  $

Multiplicative identification probabilities under independence.
\end{center}

\subsection{Independent Joint Distributions}

A joint probability distribution $p\left(  x,y\right)  $ on $X\times Y$ is
\textit{independent} if each value is the product of the marginals: $p\left(
x,y\right)  =p\left(  x\right)  p\left(  y\right)  $.

For an independent distribution, the Shannon mutual information

\begin{center}
$I\left(  x,y\right)  =\sum_{x\in X,y\in Y}p\left(  x,y\right)  \log\left(
\frac{p\left(  x,y\right)  }{p\left(  x\right)  p\left(  y\right)  }\right)  $
\end{center}

\noindent is immediately seen to be zero so we have:

\begin{center}
$H\left(  x,y\right)  =H\left(  x\right)  +H\left(  y\right)  $

Shannon entropies for independent $p\left(  x,y\right)  $.
\end{center}

For the logical mutual information, independence gives:%

\begin{align*}
m\left(  x,y\right)   &  =%
{\textstyle\sum\nolimits_{x,y}}
p\left(  x,y\right)  \left[  1-p\left(  x\right)  -p\left(  y\right)
+p\left(  x,y\right)  \right] \\
&  =%
{\textstyle\sum\nolimits_{x,y}}
p\left(  x\right)  p\left(  y\right)  \left[  1-p\left(  x\right)  -p\left(
y\right)  +p\left(  x\right)  p\left(  y\right)  \right] \\
&  =%
{\textstyle\sum\nolimits_{x}}
p\left(  x\right)  \left[  1-p\left(  x\right)  \right]
{\textstyle\sum\nolimits_{y}}
p\left(  y\right)  \left[  1-p\left(  y\right)  \right] \\
&  =h\left(  x\right)  h\left(  y\right)
\end{align*}

\begin{center}
Logical entropies for independent $p\left(  x,y\right)  $.
\end{center}

This independence condition $m\left(  x,y\right)  =h\left(  x\right)  h\left(
y\right)  $ plus the inclusion-exclusion principle $m\left(  x,y\right)
=h\left(  x\right)  +h\left(  y\right)  -h\left(  x,y\right)  $ also implies that:%

\begin{align*}
\left[  1-h\left(  x\right)  \right]  \left[  1-h\left(  y\right)  \right]
&  =1-h\left(  x\right)  -h\left(  y\right)  +h\left(  x\right)  h\left(
y\right) \\
&  =1-h\left(  x\right)  -h\left(  y\right)  +m\left(  x,y\right) \\
&  =1-h\left(  x,y\right)  \text{.}%
\end{align*}

\noindent Hence under independence, the probability of drawing the same pair
$\left(  x,y\right)  $ in two independent draws is equal to the probability of
drawing the same $x$ times the probability of drawing the same $y$.

\section{Cross-entropies and divergences}

Given two probability distributions $p=\left(  p_{1},...,p_{n}\right)  $ and
$q=\left(  q_{1},...,q_{n}\right)  $ on the same sample space $\left\{
1,...,n\right\}  $, we can again consider the drawing of a pair of points but
where the first drawing is according to $p$ and the second drawing according
to $q$. The probability that the points are distinct would be a natural and
more general notion of logical entropy that would be the:

\begin{center}
$h\left(  p\Vert q\right)  =\sum_{i}p_{i}(1-q_{i})=1-\sum_{i}p_{i}q_{i}$

\textit{Logical} \textit{cross entropy of }$p$ \textit{and} $q$
\end{center}

\noindent which is symmetric. The logical cross entropy is the same as the
logical entropy when the distributions are the same, i.e., if $p=q$, then
$h\left(  p\Vert q\right)  =h\left(  p\right)  $.

The notion of \textit{cross entropy} in Shannon entropy can be developed by
applying dit-bit connection to the logical cross entropy $\sum_{i}%
p_{i}(1-q_{i})$ to obtain:

\begin{center}
$H\left(  p\Vert q\right)  =\sum_{i}p_{i}\log\left(  \frac{1}{q_{i}}\right)  $
\end{center}

\noindent which is not symmetrical due to the asymmetric role of the
logarithm, although if $p=q$, then $H\left(  p\Vert q\right)  =H\left(
p\right)  $. Since the logical cross entropy is symmetrical, it could also be
expressed as $\sum_{i}q_{i}\left(  1-p_{i}\right)  $ which develops to the
Shannon cross entropy $H\left(  q||p\right)  =\sum_{i}q_{i}\log\left(
\frac{1}{p_{i}}\right)  $ so it might be more reasonable to use a
\textit{symmetrized cross entropy}:

\begin{center}
$H_{s}\left(  p||q\right)  =\frac{1}{2}\left[  H\left(  p||q\right)  +H\left(
q||p\right)  \right]  $.
\end{center}

The \textit{Kullback-Leibler divergence }(or \textit{relative entropy})
$D\left(  p\Vert q\right)  =\sum_{i}p_{i}\log\left(  \frac{p_{i}}{q_{i}%
}\right)  $ is defined as a measure of the distance or divergence between the
two distributions where $D\left(  p\Vert q\right)  =H\left(  p\Vert q\right)
-H\left(  p\right)  $. A basic result is the:

\begin{center}
$D\left(  p\Vert q\right)  \geq0$ with equality if and only if $p=q$

\textit{Information inequality} \cite[p. 26]{cover:eit}.
\end{center}

Given two partitions $\pi$ and $\sigma$, the inequality $I\left(  \pi
,\sigma\right)  \geq0$ is obtained by applying the information inequality to
the two distributions $\left\{  p_{B\cap C}\right\}  $ and $\left\{
p_{B}p_{C}\right\}  $ on the sample space $\left\{  \left(  B,C\right)
:B\in\pi,C\in\sigma\right\}  =\pi\times\sigma$:

\begin{center}
$I\left(  \pi,\sigma\right)  =\sum_{B,C}p_{B\cap C}\log\left(  \frac{p_{B\cap
C}}{p_{B}p_{C}}\right)  =D\left(  \left\{  p_{B\cap C}\right\}  \Vert\left\{
p_{B}p_{C}\right\}  \right)  \geq0$

with equality iff independence.
\end{center}

\noindent In the same manner, we have for the joint distribution $p\left(
x,y\right)  $:

\begin{center}
$I\left(  x,y\right)  =D\left(  p\left(  x,y\right)  ||p\left(  x\right)
p\left(  y\right)  \right)  \geq0$

with equality iff independence.
\end{center}

The \textit{symmetrized Kullback-Leibler divergence} is:

\begin{center}
$D_{s}(p||q)=\frac{1}{2}\left[  D\left(  p||q\right)  +D\left(  q||p\right)
\right]  =H_{s}\left(  p||q\right)  -\left[  \frac{H\left(  p\right)
+H\left(  q\right)  }{2}\right]  $.
\end{center}

But starting afresh, one might ask: \textquotedblleft What is the natural
measure of the difference or distance between two probability distributions
$p=\left(  p_{1},...,p_{n}\right)  $ and $q=\left(  q_{1},...,q_{n}\right)  $
that would always be non-negative, and would be zero if and only if they are
equal?\textquotedblright\ The (Euclidean) distance between the two points in $%
\mathbb{R}
^{n}$ would seem to be the \textquotedblleft logical\textquotedblright%
\ answer---so we take that distance (squared with a scale factor) as the
definition of the:

\begin{center}
$d\left(  p\Vert q\right)  =$ $\frac{1}{2}\sum_{i}\left(  p_{i}-q_{i}\right)
^{2}$

\textit{Logical divergence} (or \textit{logical} \textit{relative
entropy})\footnote{In \cite{ell:countdits}, this definition was given without
the useful scale factor of $1/2$.}
\end{center}

\noindent which is symmetric and we trivially have:

\begin{center}
$d\left(  p||q\right)  \geq0$ with equality iff $p=q$

Logical information inequality.
\end{center}

We have component-wise:

\begin{center}
$0\leq\left(  p_{i}-q_{i}\right)  ^{2}=p_{i}^{2}-2p_{i}q_{i}+q_{i}%
^{2}=2\left[  \frac{1}{n}-p_{i}q_{i}\right]  -\left[  \frac{1}{n}-p_{i}%
^{2}\right]  -\left[  \frac{1}{n}-q_{i}^{2}\right]  $
\end{center}

\noindent so that taking the sum for $i=1,...,n$ gives:

\begin{center}%
\begin{align*}
d\left(  p\Vert q\right)   &  =\frac{1}{2}%
{\textstyle\sum\nolimits_{i}}
\left(  p_{i}-q_{i}\right)  ^{2}\\
&  =\left[  1-%
{\textstyle\sum\nolimits_{i}}
p_{i}q_{i}\right]  -\frac{1}{2}\left[  \left(  1-%
{\textstyle\sum\nolimits_{i}}
p_{i}^{2}\right)  +\left(  1-%
{\textstyle\sum\nolimits_{i}}
q_{i}^{2}\right)  \right] \\
&  =h\left(  p\Vert q\right)  -\frac{h\left(  p\right)  +h\left(  q\right)
}{2}\text{.}%
\end{align*}

Logical divergence = \textit{Jensen difference} \cite[p. 25]{rao:div} between
probability distributions.
\end{center}

\noindent Then the information inequality implies that the logical
cross-entropy is greater than or equal to the average of the logical entropies:

\begin{center}
$h\left(  p||q\right)  \geq\frac{h\left(  p\right)  +h\left(  q\right)  }{2}$
with equality iff $p=q$.
\end{center}

The half-and-half probability distribution $\frac{p+q}{2}$ that mixes $p$ and
$q$ has the logical entropy of

\begin{center}
$h\left(  \frac{p+q}{2}\right)  =\frac{h\left(  p\Vert q\right)  }{2}%
+\frac{h\left(  p\right)  +h\left(  q\right)  }{4}=\frac{1}{2}\left[  h\left(
p||q\right)  +\frac{h\left(  p\right)  +h\left(  q\right)  }{2}\right]  $
\end{center}

so that:

\begin{center}
$h(p||q)\geq h\left(  \frac{p+q}{2}\right)  \geq\frac{h\left(  p\right)
+h\left(  q\right)  }{2}$ with equality iff $p=q$.

Mixing different $p$ and $q$ increases logical entropy.
\end{center}

The logical divergence can be expressed as:

\begin{center}
$d\left(  p\Vert q\right)  =\frac{1}{2}\left[  \sum_{i}p_{i}\left(
1-q_{i}\right)  +\sum_{i}q_{i}\left(  1-p_{i}\right)  \right]  -\frac{1}%
{2}\left[  \left(  \sum_{i}p_{i}\left(  1-p_{i}\right)  \right)  +\left(
\sum_{i}q_{i}\left(  1-q_{i}\right)  \right)  \right]  $
\end{center}

\noindent that develops via the dit-bit connection to:

$\frac{1}{2}\left[  \sum_{i}p_{i}\log\left(  \frac{1}{q_{i}}\right)  +\sum
_{i}q_{i}\log\left(  \frac{1}{p_{i}}\right)  -\sum_{i}p_{i}\log\left(
\frac{1}{p_{i}}\right)  -\sum_{i}q_{i}\log\left(  \frac{1}{q_{i}}\right)
\right]  $

$=\frac{1}{2}\left[  \sum_{i}p_{i}\log\left(  \frac{p_{i}}{q_{i}}\right)
+\sum_{i}q_{i}\log\left(  \frac{q_{i}}{p_{i}}\right)  \right]  =\frac{1}%
{2}\left[  D\left(  p||q\right)  +D\left(  q||p\right)  \right]  $

$=D_{s}\left(  p||q\right)  $.

\noindent Thus the logical divergence $d\left(  p||q\right)  $ develops via
the dit-bit connection to the symmetrized version of the Kullback-Leibler divergence.

\section{Summary and concluding remarks}

The following table summarizes the concepts for the Shannon and logical
entropies. We use the case of probability distributions rather than
partitions, and we use the abbreviations $p_{xy}=p(x,y)$, $p_{x}=p(x)$, and
$p_{y}=p\left(  y\right)  $.

\begin{center}%
\begin{tabular}
[c]{l|c|c|}\cline{2-3}
& $\text{Shannon Entropy}$ & $\text{Logical Entropy}$\\\hline\hline
\multicolumn{1}{|l|}{{\small Entropy}} & ${\small H(p)=}\sum{\small p}_{i}%
\log\left(  1/p_{i}\right)  $ & ${\small h}\left(  p\right)  {\small =}%
\sum{\small p}_{i}\left(  1-p_{i}\right)  $\\\hline
\multicolumn{1}{|l|}{{\small Mutual Info.}} & ${\small I(x,y)=H}\left(
x\right)  {\small +H}\left(  y\right)  {\small -H}\left(  x,y\right)  $ &
${\small m}\left(  x,y\right)  {\small =h}\left(  x\right)  {\small +h}\left(
y\right)  {\small -h}\left(  x,y\right)  $\\\hline
\multicolumn{1}{|l|}{{\small Independence}} & ${\small I}\left(  x,y\right)
{\small =0}$ & ${\small m}\left(  x,y\right)  {\small =h}\left(  x\right)
{\small h}\left(  y\right)  $\\\hline
\multicolumn{1}{|l|}{{\small Indep. Relations}} & ${\small H}\left(
x,y\right)  {\small =H}\left(  x\right)  {\small +H}\left(  y\right)  $ &
{\small \ }${\small 1-h}\left(  x,y\right)  {\small =}\left[  1-h\left(
x\right)  \right]  \left[  1-h\left(  y\right)  \right]  $\\\hline
\multicolumn{1}{|l|}{{\small Cond. entropy}} & {\small \ }${\small H}\left(
x|y\right)  {\small =}\sum_{x,y}{\small p}_{xy}\log\left(  \frac{p_{y}}%
{p_{xy}}\right)  $ & ${\small h}\left(  x|y\right)  {\small =}\sum
_{x,y}{\small p}_{xy}\left[  \left(  p_{y}-p_{xy}\right)  \right]  $\\\hline
\multicolumn{1}{|l|}{{\small Relationships}} & ${\small H}\left(  x|y\right)
{\small =H}\left(  x,y\right)  {\small -H}\left(  y\right)  $ & ${\small h}%
\left(  x|y\right)  {\small =h}\left(  x,y\right)  {\small -h}\left(
y\right)  $\\\hline
\multicolumn{1}{|l|}{{\small Cross entropy}} & ${\small H}\left(  p\Vert
q\right)  {\small =}\sum{\small p}_{i}\log\left(  1/q_{i}\right)  $ &
${\small h}\left(  p\Vert q\right)  {\small =}\sum{\small p}_{i}\left(
1-q_{i}\right)  $\\\hline
\multicolumn{1}{|l|}{{\small Divergence}} & ${\small D}\left(  p\Vert
q\right)  {\small =}\sum_{i}{\small p}_{i}\log\left(  \frac{p_{i}}{q_{i}%
}\right)  $ & {\small \ }${\small d}\left(  p||q\right)  {\small =}\frac{1}%
{2}\sum_{i}\left(  p_{i}-q_{i}\right)  ^{2}$\\\hline
\multicolumn{1}{|l|}{{\small Relationships}} & ${\small D}\left(  p\Vert
q\right)  {\small =H}\left(  p\Vert q\right)  {\small -H}\left(  p\right)  $ &
${\small d}\left(  p\Vert q\right)  {\small =h}\left(  p\Vert q\right)
{\small -}\left[  {\small h}\left(  p\right)  {\small +h}\left(  q\right)
\right]  /2$\\\hline
\multicolumn{1}{|l|}{{\small Info. Inequality}} & ${\small D}\left(  p\Vert
q\right)  {\small \geq0}\text{ with }{\small =}\text{ iff }{\small p=q}$ &
${\small d}\left(  p\Vert q\right)  {\small \geq0}\text{ with }{\small =}%
\text{ iff }{\small p=q}$\\\hline
\end{tabular}

Table of comparisons between Shannon and logical entropies
\end{center}

The above table shows many of the same relationships holding between the
various forms of the logical and Shannon entropies due ultimately to the
dit-bit connection. The dit-bit connection between the two notions of entropy
is based on them being two different measures of the "amount of
information-as-distinctions," the dit-count being the normalized count of the
distinctions and the bit-count being the number of binary partitions required
(on average) to make the distinctions.

Logical entropies arise naturally as the normalized counting measure for
partition logic just as probabilities arise as the normalized counting measure
for subset logic, where the two logics are dual to one another. All the forms
of logical entropy have simple interpretations as the probabilities of
distinctions. Shannon entropy is a higher-level and more refined notion
adapted to the theory of communications and coding where it can be interpreted
as the average number of bits necessary per letter to code the messages, i.e.,
the average number of binary partitions necessary per letter to distinguish
the messages.

\end{document}